\documentclass[usenatbib,onecolumn]{mn2e}

\usepackage{amsmath, mathrsfs, amssymb}
\usepackage{graphicx}
\usepackage{hyperref}
\usepackage{rotating}
\usepackage{longtable,lscape}
\usepackage{morefloats}
\usepackage{color}
\usepackage[usenames,dvipsnames,svgnames,table]{xcolor}
\usepackage{caption}
\usepackage{verbatim}
\usepackage{color}
\usepackage{bm}
\usepackage[T1]{fontenc}
\usepackage{aecompl}

\title[Instability in strongly
magnetized accretion discs]{Instability in strongly
magnetized accretion discs: A global perspective}

\author[]{}
\author[Das, Begelman \& Lesur]{Upasana~Das$^{1}\thanks{E-mail: upasana.das@jila.colorado.edu}$,
Mitchell~C.~Begelman$^{1,2}$ and Geoffroy Lesur$^{1,3,4}\thanks{JILA Visiting Fellow}$ \\
$^{1}${JILA, University of Colorado and National Institute of Standards and Technology, 440 UCB, Boulder,
CO 80309-0440, USA.} \\
$^{2}${Department of Astrophysical and Planetary Sciences, University of Colorado, 391 UCB, Boulder, CO 80309-0391, USA.} \\
$^{3}${Univ. Grenoble Alpes, IPAG, 38000 Grenoble, France.}  \\
$^{4}${CNRS, IPAG, F-38000 Grenoble, France.}}

\begin{document}
\label{firstpage}
\maketitle

\begin{abstract}
We examine the properties of strongly magnetized accretion discs in a global framework,
with particular focus on the evolution of magnetohydrodynamic instabilities
such as the magnetorotational instability
(MRI). Work by Pessah \& Psaltis showed that MRI is stabilized beyond a critical toroidal
field in compressible, differentially rotating flows and, also, reported the
appearance of two new instabilities beyond this field. Their results stemmed
from considering geometric curvature effects due to the suprathermal background
toroidal field, which had been previously ignored in weak-field studies. However,
their calculations were performed under the local approximation, which poses the danger
of introducing spurious behavior due to the introduction of global geometric terms
in an otherwise local framework. In order to avoid this, we perform a global eigenvalue
analysis of the linearized MHD equations in cylindrical geometry. We confirm that MRI
indeed tends to be highly suppressed when the background toroidal field attains the Pessah-Psaltis limit.
We also observe the appearance of two new instabilities that emerge in the presence of
highly suprathermal toroidal fields. These results were additionally verified using
numerical simulations in PLUTO. There are, however, certain differences between the
the local and global results, especially in the vertical wavenumber
occupancies of the various instabilities, which we discuss in detail. We also study
the global eigenfunctions of the most unstable modes in the suprathermal
regime, which are inaccessible in the local analysis. Overall, our findings emphasize the
necessity of a global treatment for accurately modeling strongly magnetized accretion discs.
\end{abstract}

\begin{keywords}
accretion, accretion discs - instabilities - magnetohydrodynamics (MHD)
\vspace{-6mm}
\end{keywords}

\section{Introduction}
\label{sec_intro}

The magnetorotational instability
\citep[MRI;][]{1991ApJ...376..214B,1998RvMP...70....1B}, which
occurs in differentially rotating plasmas threaded by a weak magnetic field, is
believed to explain the long-standing puzzle of the origin of turbulence and angular
momentum transport in hydrodynamically stable accretion discs around compact objects.
Most of the accretion disc studies
so far have focused on the weak-field regime corresponding to plasma-$\beta \gg 1$, where
$\beta \equiv P/P_{B}$, $P$ and $P_B$ being the gas and the (total) magnetic pressures,
respectively. This is probably due to the apparent
difficulty in explaining the generation and sustenance of a strong magnetic field therein.
However, there is growing evidence, from
both numerical simulations and observations, of accretion flows having suprathermal magnetic fields
($\beta <1$).

For instance, \citet{2000ApJ...534..398M} carried out three-dimensional
vertically stratified shearing box simulations and studied the evolution of
initially subthermal magnetic fields in the disc midplane having different
geometries, namely, purely toroidal, zero net vertical flux and a uniform
vertical field. They reported the formation of
magnetically dominated ($\beta <1$) coronae
above 2 scale heights from the disc midplane for the first two cases, due to the
amplification of the magnetic field by MRI and its subsequent rise due to magnetic buoyancy.
They also observed the formation of a large-scale field structure in both the disc and
the corona, which was dominated by the toroidal field component.
For the uniform vertical field case they found both the corona and the disc to be
magnetically dominated. However, this analysis was limited to the
linear regime only, due to numerical difficulties arising from the generation
of the disruptive MRI channel flow and, hence, is inconclusive.
\citet{2013ApJ...767...30B}  carefully avoided the generation of the MRI channel
flow in their vertically stratified shearing box simulations, which enabled
them to reach very high initial net vertical fluxes. They found the entire
disc to be magnetically dominated for initial vertical fields
stronger than the threshold value of $\beta \sim 10^3$ at the midplane.
More importantly, they observed the generation of a large-scale
toroidal field that became suprathermal for such a strong initial vertical flux
--- the resulting time-averaged $\beta$ being in the range $0.1-1$ (see, e.g., Figure 3 of
\citealt{2013ApJ...767...30B}) --- a result that is of particular relevance to this work.
More recently,
\citet{2016MNRAS.460.3488S,2016MNRAS.457..857S}
carried out comprehensive stratified shearing box simulations by extending the work of
\citet{2013ApJ...767...30B}, in order to further explore
the properties of MRI turbulence and dynamo activity in strongly
magnetized accretion discs. Their results indicate the necessity of a net
initial vertical flux in order to yield a
magnetically dominated steady state accretion disc, which is required
by the dynamo to continuously regenerate and sustain
the buoyantly escaping toroidal field (also verified by the
global disc simulations of \citealt{2017MNRAS.467.1838F}).
An important aspect of a strongly magnetized disc is the
possibility of magnetocentrifugally driven outflows.
\citet{2013A&A...550A..61L} studied stratified shearing box simulations of an accretion disc
having a strong poloidal field with  $\beta \sim 10$, and found a comprehensible
link between MRI and the generation of disc winds (also, see
\citealt{2012A&A...548A..76M,2013A&A...552A..71F,2013ApJ...767...30B,2016MNRAS.463.3096R}).
There have also been observations of strongly magnetized winds in
black hole binaries such as GRO J1655-40 \citep{2006Natur.441..953M} and
GRS 1915+105 \citep{2016ApJ...821L...9M},
indicative of the probable  existence of an
underlying strongly magnetized disc from which the winds are launched.

Strongly magnetized accretion discs have been theoretically
shown to be viscously and thermally stable
\citep{2007MNRAS.375.1070B,2009ApJ...697...16O,2016MNRAS.459.4397S},
a result which might
finally lead to a fruitful resolution of the mismatch between
observations and theoretical predictions of the standard $\alpha$-disc model
\citep{1973A&A....24..337S,1976MNRAS.175..613S,1974ApJ...187L...1L}.
Magnetically dominated discs
are also less dense than their weakly
magnetized counterparts, which makes them more stable against fragmentation
under self-gravity \citep*{2003A&A...407..403P}. This in turn can prevent
the clumping of the infalling gas and aid the fueling of active galactic nuclei and
the growth of supermassive black hole seeds
\citep*{2007MNRAS.375.1070B,2012ApJ...758..103G}.
Strongly magnetized discs can also explain the larger than expected vertical scale heights,
inflow speeds and color temperatures inferred
for accretion discs in systems such as cataclysmic variables and some X-ray binaries \citep{2007MNRAS.375.1070B},
as well as several key aspects of
X-ray binary state transitions, which are otherwise
unresolved in the standard disc theory \citep*{2015ApJ...809..118B}.

Given the likely importance of strong magnetic fields in accretion
discs, it is essential to determine whether and how the instabilities like MRI
are affected in this regime.
\citet[][hereafter PP05]{2005ApJ...628..879P}
showed by means of a local linear
stability analysis that suprathermal toroidal magnetic fields (in the presence of a highly sub-thermal
poloidal field) have a profound effect on the stability of
large vertical wavenumber, axisymmetric perturbations, which correspond to the most unstable modes
of conventional MRI. They demonstrated that this was mainly
due to the important roles played by both the
curvature of the toroidal field lines and compressibility when $\beta < 1$ ---
either or both effects being largely
ignored in weak-field studies
\citep[see e.g.][and PP05 for a detailed comparison]{1991ApJ...376..214B,
1992ApJ...392..662B,1998RvMP...70....1B,1994ApJ...421..163B,1995ApJ...453..380B}.
Interestingly,
their study identified a critical toroidal Alfv\'{e}n speed for purely Keplerian flows, beyond
which MRI starts to get
stabilized at small vertical wavenumbers (i.e., $v_{ A \phi}^{PP1} = \sqrt{v_K c_s}$, where $v_{A\phi}^{ PP1}$,
$v_K$ and $c_s$ are the critical toroidal Alfv\'{e}n velocity, Keplerian
velocity and sound speed, respectively) and
is eventually completely suppressed, across the entire range of allowed vertical wavenumbers,
at a slightly higher critical
value (i.e., $v_{A \phi}^{PP2} =\sqrt{2 v_K c_s} $).
Additionally, they reported the emergence of two new suprathermal
instabilities, beyond $v_{A \phi}^{PP2}$, that occupy different wavenumber regimes.
Such an upper limit on the magnetic field strength for
MRI to operate is quite appealing as, if correct, it not only helps
constrain the theory better, but also provides testable predictions. However,
since PP05 carried out their analysis under the local approximation,
one cannot be confident regarding the robustness of their findings. This is
because incorporating global curvature terms in a local framework often leads to spurious outcomes.

Our main aim in this work is to revisit the analysis by PP05 and reassess
their main results in a global framework, which is a necessary step before extending the
model to add more complex physics.
In order to do so, we solve the global, linear eigenvalue problem for a compressible,
axisymmetric, magnetized fluid in a cylindrical disc geometry.
This also allows us to systematically compare the results in the two
formalisms \citep[also, see][for the connection between local
and global weak-field MRI modes]{1994ApJ...434..206C, 2015MNRAS.453.3257L}.
In our analysis, we find that MRI indeed tends to be highly
suppressed for sufficiently suprathermal toroidal
background fields.
However, unlike PP05, we observe only a partial reduction
in the MRI growth rate at small vertical
wavenumbers when the background toroidal Alfv\'{e}n velocity exceeds $v_{A \phi}^{PP1}$.
In fact, as long as MRI operates in the global analysis,
it spans the entire range of allowed vertical wavenumbers.
We also observe the appearance of two new instabilities, as predicted by PP05, when the
background toroidal Alfv\'{e}n velocity exceeds $v_{A \phi}^{PP2}$.
However, the global
results exhibit a very different
variation of growth rate as a function of vertical
wavenumber for these two instabilities. Overall, it appears
that the local analysis predicts
the maximum possible growth rates of the
various instabilities in the suprathermal regime reasonably well,
but falls short in estimating the range of vertical wavenumbers occupied by
them.
Nevertheless, this is an important confirmation, since a local analysis, if valid to a
reasonable degree, gives us a much better
understanding of the underlying stability criteria.
We further carry out a small set of simulations using
the finite volume code PLUTO \citep{2007ApJS..170..228M}, 
which corroborates the results from our global eigenvalue analysis.
The current work is the first in a series of explorations, which
include additional effects such as non-axisymmetry, and radial and vertical stratification
in a strongly magnetized disc.

In this context, we mention that \citet{1995ApJ...453..697C} carried out one of the earlier
global, linear stability analyses
of a differentially rotating flow to study the effect of a strong toroidal magnetic field.
They also noted the progressive stabilization of MRI with increasing toroidal field strength.
However, their work considered only {\it incompressible} flows (additionally having radial stratification)
and, hence, some of their conclusions
differ from those of PP05.
 For instance, the critical toroidal Alfv\'{e}n speed for complete MRI stabilization according to
\citet{1995ApJ...453..697C}
is given by the rotational speed of the flow, which is justified as it is
the only velocity scale in the absence of compressibility.
They also
reported the appearance of a new instability called the large field instability (LFI),
which we will discuss later in \S \ref{subsec_supra}.
Thus, compressibility along with magnetic curvature seems to play an important role
in determining the stability of a strongly magnetized accretion flow.

We also mention here the linear
stability analysis of an axisymmetric accretion flow carried out
by \citet[][hereafter TP96]{1996MNRAS.279..767T}
in the presence of a purely toroidal, but subthermal magnetic field, and vertical and radial stratification.
They performed local
as well as global analyses and found the results of the two cases
to be in good agreement with each other.
They observed mainly two kinds of instabilities for
the extreme limits of their local dispersion relation.
When the ratio of radial to vertical wavenumber was large,
a Parker type instability was observed, driven by vertical magnetic buoyancy.
On the other hand, when the ratio of radial to vertical wavenumber was small, a shear driven instability
prevailed, even in the absence of stratification.
They also noted that both kinds of instabilities were highly localized in the radial coordinate.
The local dispersion
relation derived in this work generalizes the one derived by
TP96 by including a uniform background poloidal field,
although it does not include vertical stratification (see Appendix
\ref{sec_limitcases}).
However, suprathermal background toroidal fields, together
with a subthermal poloidal field and no vertical stratification, yield
very different instabilities in our work. We still obtain shear driven
modes but do not observe  magnetic vertical buoyancy driven modes (due to
the absence of vertical stratification in our work). We instead observe
the appearance of radial buoyancy driven modes, due to
the radial tension force
from the suprathermal toroidal field \citep[also, see][in a subthermal context]{2000ApJ...540..372K},
which we elaborate below.


This paper is organized as follows. In \S \ref{eqs_sec}, we lay out the linearized MHD
equations that form the basis of our analysis. In \S \ref{sec_local}, we
focus on developing a self-consistent local theory and
present the calculations leading to a generic local dispersion
relation, which includes the effects of magnetic curvature,
compressibility, non-axisymmetry and background radial gradients.
We also obtain the PP05 limit of our local dispersion relation in order
to compare with the global analysis.
In \S \ref{sec_global}, we describe in detail the numerical set-up and normalization scheme
used for our global eigenvalue analysis.
In \S \ref{sec_global_solns}, we present the solutions of the global eigenvalue
problem. We first recall the stability criteria from local theory and, then,
systematically analyze the global properties, including the global eigenfunctions,
of the {\it most} unstable modes of the instabilities in the suprathermal regime.
In \S \ref{sec_pluto}, we present the results of a small set of
numerical simulations performed using PLUTO and compare them with our
global eigenvalue analysis.
We conclude
in \S \ref{sec_conc}, by highlighting our main results and discussing some of their
implications, which readers may refer to at any point for a brief summary
of this work.

\section{Linearized MHD equations of motion}
\label{eqs_sec}

We begin by recalling the ideal MHD equations characterizing a magnetized, compressible accretion flow

\begin{gather}
\frac{\partial \rho}{\partial t} + \nabla \cdot (\rho \mathbf{v}) = 0 ~, \label{mhd1} \\
\frac{\partial \mathbf{v}}{\partial t} + (\mathbf{v} \cdot \nabla ) \mathbf{v} + \nabla \Phi +
\frac{1}{\rho}\nabla \left(P + \frac{\mathbf{B}^2} {8\pi}\right) - \frac{1}{4\pi \rho}(\mathbf{B}\cdot \nabla)\mathbf{B} = 0 ~,
\label{mhd2}\\
\frac{\partial \mathbf{B}}{\partial t} - \nabla \times (\mathbf{v}\times  \mathbf{B}) = 0 ~,
\label{mhd3}
\end{gather}
where $\rho$ is the density, $P$ the gas pressure,  $\Phi$ the gravitational potential
of the central object,
$\bf{B}$ the magnetic field and $\bf{v}$ the fluid velocity. We adopt a cylindrical
co-ordinate system ($r, \phi, z$) and an axisymmetric background
having ${\bf v} = (0, v_{0 \phi}(r), 0)$ and $ {\bf B} =(0, B_{0\phi}(r), B_{0z})$, where $B_{0z}$ is a constant
and the other flow variables are assumed to depend on $r$ only. We neglect
vertical stratification and vertical gravity, and assume the
gravitational potential to be cylindrical such that $\Phi = -GM/r$.
The above equations are also supplemented by an equation of state, $P = P(\rho)$.
We furthermore neglect self gravity of the disc, as well as any dissipative processes.


In order to carry out a linear stability analysis, we perturb equations (\ref{mhd1})-(\ref{mhd3})
in Eulerian coordinates
and retain only the perturbed amplitudes of linear order such that
\begin{gather}
\rho = \rho_0 + \rho_1 ~, \\
P = P_0 + P_1 ~,\\
{\bf v} = {\bf v_0} + {\bf v_1} ~,\\
{\bf B} = {\bf B_0} + {\bf B_1} ~.
\end{gather}
Note that whereas the vertical equilibrium is trivial in the absence of any vertical stratification,
the radial equilibrium is derived from equation (\ref{mhd2}) as
\begin{equation}
\Omega^2 = \Omega_K^2 + \frac{v_{A\phi}^2} {r^2} \left(1 +  \frac{\partial \ln B_{0\phi}}{ \partial \ln r} \right)
+ \frac{1}{r\rho_0}\partial_r P_0 ~,
\label{eq_radeqbm}
\end{equation}
where, $\Omega = v_{0\phi}/r$ is the background angular velocity, $\Omega_K = \sqrt{(GM/r^3)}$ the Keplerian
rotation frequency and $v_{A\phi} = B_{0\phi}/ \sqrt{(4\pi\rho_0)}$ the toroidal Alfv\'{e}n velocity.
Thus, we see that such a disc may deviate significantly from purely Keplerian
rotation, especially if it is strongly magnetized.
However, Keplerian rotation is recovered in the absence of background radial
gradients and for $v_{A\phi} \ll v_K$, where $v_K = \sqrt{(GM/r)}$ is
the Keplerian velocity.

We now write below the complete set of the linearized MHD equations:
\begin{equation}
\partial_t  \rho_1 + \rho_0 \biggl[ \frac{1}{r} \partial_r (r  v_{1r}) + \frac{1}{r} \partial_\phi  v_{1\phi} +
\partial_z  v_{1z} \biggr] + \Omega \partial_\phi \rho_1 + v_{1r} \partial_r \rho_0 = 0 ~,
\label{eq_cont}
\end{equation}
\begin{align}
\rho_0 \biggl[ \partial_t v_{1r} +  \Omega \partial_\phi v_{1r} - 2\Omega v_{1\phi} \biggr] + \partial_r P_1
+ \frac{1}{4 \pi} \biggl[ B_{0\phi} \partial_r B_{1 \phi} &+ B_{1\phi}\partial_r B_{0\phi}
-  \frac{B_{0\phi}}{r} \partial_\phi B_{1r}
- B_{0z} \partial_z B_{1r} + B_{0z} \partial_r B_{1z}   \nonumber \\ &+   \frac{2 B_{0\phi} B_{1\phi}}{r}   \biggr]
- \frac{\rho_1}{\rho_0} \biggl[  \frac{B_{0\phi}^2}{4 \pi r} + \partial_r P_0 +
\frac{B_{0\phi}}{4\pi} \partial_r B_{0\phi}  \biggr] = 0 ~,
\label{eq_radmom}
\end{align}
\begin{equation}
\rho_0 \biggl[ \partial_t v_{1\phi} + v_{1r} (r \partial_r \Omega) +
2\Omega v_{1r} + \Omega \partial_\phi v_{1\phi}  \biggr]
+ \frac{1}{r} \partial_\phi P_1 - \frac{1}{4\pi} \biggl[ B_{1r}\partial_r B_{0\phi} + B_{0z} \partial_z B_{1\phi}
- \frac{B_{0z}}{r} \partial_\phi B_{1z} + \frac{B_{0\phi} B_{1r}}{r}    \biggr]  = 0 ~,
\label{eq_phimom}
\end{equation}
\begin{equation}
\rho_0 \biggl[ \partial_t v_{1z} + \Omega \partial_\phi v_{1z} \biggr]  + \partial_z P_1 +
\frac{1}{4\pi} \biggl[ B_{0\phi} \partial_z  B_{1\phi} - \frac{B_{0\phi}}{r}\partial_\phi B_{1z}  \biggr]  = 0~,
\label{eq_zmom}
\end{equation}
\begin{equation}
\partial_t B_{1r} + \Omega \partial_\phi B_{1r} - \frac{B_{0\phi}}{r} \partial_\phi v_{1r} - B_{0z} \partial_z v_{1r}  = 0~,
\label{eq_br}
\end{equation}
\begin{equation}
\partial_t B_{1\phi} + v_{1r} \partial_r B_{0\phi} + \Omega\partial_\phi B_{1\phi} - B_{1r}( r\partial_r \Omega)
- B_{0z} \partial_z v_{1\phi} + B_{0\phi} \partial_z v_{1z} + B_{0\phi} \partial_r v_{1r} = 0~,
\label{eq_bphi}
\end{equation}
\begin{equation}
\partial_t B_{1z} + \Omega \partial_\phi B_{1z} - \frac{B_{0\phi}}{r} \partial_\phi v_{1z} +
\frac{B_{0z}}{r} \partial_\phi v_{1\phi} + \frac{B_{0z}}{r} \partial_r (r v_{1r}) = 0  ~.
\label{eq_bz}
\end{equation}
Note that the above set of equations is quite generic, encompassing non-axisymmetric perturbations,
background radial gradients, compressibility and magnetic curvature effects, and is generally applicable
for any magnetic field strength.
We can now use the same linearized system of equations to derive a local dispersion relation, as well as
solve the global eigenvalue problem.

\section{Local analysis}
\label{sec_local}

In this section, we present a systematic derivation of the local dispersion relation
from the set of equations (\ref{eq_cont})-(\ref{eq_bz}), which differs from the approach
employed by PP05.
We apply a more physically motivated WKB formalism and, also, account for
the radial dependence of the normal modes in the system to the requisite order
(as both the amplitude and radial wavenumber of the modes should in general be functions of
radius when there is a radially varying background).
We explicitly include the magnetic tension in the background flow such that it
is no longer purely Keplerian in the strongly magnetized limit, in order to make the
global analysis self-consistent.
Additionally, we obtain a more generic
dispersion relation containing the effects of non-axisymmetry and background radial gradients.
We discuss some of the well known limiting cases of our local dispersion relation in
Appendix \ref{sec_limitcases}.


\subsection{Analytical approximation scheme}
\label{sec_approx}

Before proceeding further, we first apply certain physically motivated assumptions
that simplify the problem
and make the solutions analytically tractable. We also define some new variables for a more
compact visualization during the analysis in the present section, which are listed in Table \ref{tab_notation}.

\begin{itemize}


\item We assume that the perturbations are adiabatic such that the energy equation is given by
\begin{equation}
\dot P_1 - c_s^2 \dot \rho_1 =   - (\gamma P_0) ({\bf v_1} \cdot {\bf S}) ~,
\end{equation}
where the over-dot represents time derivative, $c_s = \sqrt{\gamma P/\rho}$ is the local sound speed,
$\gamma$ the adiabatic index
and ${\bf S}$ the Schwarzschild discriminant vector or non-adiabacity of the fluid given by
\begin{equation}
{\bf S} = \nabla \biggl( \ln \frac{P_0^{1/\gamma}}{\rho_0}  \biggr) ~.
\label{eq_adiab}
\end{equation}
For simplicity, we will assume that the background is adiabatic so that ${\bf S} = 0$
(see Appendix \ref{sec_noad_cor} for a more general case) and we simply have
\begin{equation}
P_1 = \frac{ \gamma P_0}{ \rho_0} \rho_1 =   c_s^2 \rho_1 ~.
\label{eq_pert_eos}
\end{equation}

\item We assume the perturbed variables to have the form
\begin{equation}
f_i = f_i(r) \exp i( m \phi + k_z z -  \omega t)
\label{eq_mode1}
\end{equation}
such that invoking the WKB approximation we can write
\begin{equation}
\partial_r f_i = i l f_i + g_i(r) ~, ~~~~ g_i = {\cal O}(\frac{1}{r}) ~,
\label{eq_mode2}
\end{equation}
where $\omega$ is the modal frequency, $t$ the time, $l$, $m$ and $k_z$
the radial, azimuthal and vertical wavenumbers (all constants)
respectively.  We focus on large wavenumber modes such that $lr, k_zr \gg 1$
and, also, $g_i(r)$ is assumed to be small in the WKB sense such that
$|g_i(r)| \ll lf_i$.
We note here that while considering radial derivatives of the perturbations,
PP05 considered only the $i l f_i$ terms in equation (\ref{eq_mode2})
and, as a result, our conclusions differ from theirs in the large $l/k_z$ limit,
as discussed below.

\item We furthermore follow the
orderings stated in \citet{1998ApJ...493..291B} such that
\begin{gather}
k_z^2 r^2 \gg 1 + m^2 ~, ~~~~ k_z = {\mathcal O}(l) ~,\\
|\omega^2| \ll  k_z^2 c_s^2 ~, ~~~~ k_z v_{1z} + l v_{1r} = \mathcal{O} \biggl(\frac{v_{1r}}{r} \biggr)~,
\label{fast_neg} \\
B_{0z} \ll B_{0\phi} ~, ~~~~ B_{0z}l, B_{0z} k_z = {\mathcal O} \biggl(\frac{B_{0\phi}}{r}\biggr)~,
\end{gather}
which further allows us to write
\begin{equation}
v_{1z}  \approx -\frac{l}{k_z}v_{1r}   ~~~{\rm and}~~~ B_{1z} \approx - \frac{l}{k_z}B_{1r} ~.
\label{v1z_B1z}
\end{equation}

\item Note that the equations (\ref{eq_cont})-(\ref{eq_bz}) would yield a dispersion relation
that is a sixth-degree polynomial in $\omega$, whose six solutions correspond to
two fast magnetosonic modes, two Alfv\'{e}n modes and two slow magnetosonic modes.
However, the fast magnetosonic modes
lie well separated from the Alfv\'{e}n and slow magnetosonic modes in the $\omega$-$k_z$ space
(as we shall indeed see later) and, hence, can be neglected in our analysis with the aim
of simplifying the derivation and obtaining some useful analytical insights. We do so by
assuming condition (\ref{fast_neg}), which can be interpreted as
neglecting the acceleration term in the vertical force balance equation. This
eventually yields a reduced fourth-degree dispersion relation in $\omega$.

\end{itemize}


\begin{table*}
\centering
\renewcommand{\arraystretch}{2}
\caption{Summary of the dimensionless parameters introduced (and to be used only)
in \S \ref{sec_local}, and Appendices \ref{sec_noad_cor} and \ref{sec_limitcases}
of this work.}
 \begin{tabular}{|cc|cc|}
 \hline
  $\eta = k_z r \frac{B_{0z}}{B_{0\phi}}$ &  & ${\bf b} = \frac{{\bf B_1}}{B_{0\phi}}$ & \\
$\hat{B}_\phi=\frac{\partial \ln B_{0\phi}}{\partial \ln r}$ &
& $q = - \frac{\partial \ln \Omega}{\partial \ln r}$ & shear parameter \\
 $\beta = \frac{8\pi P_0}{B_{0\phi}^2}$ & plasma-beta & $n = m +\eta$ &\\
 ${\bf u} = \biggl(\frac{\rho_0}{\gamma P_0}\biggr)^{1/2} {\bf v_1} \equiv \frac{1}{c_s} {\bf v_1}$
 & & $x= \frac{\gamma \beta}{2}$  & \\
  $\{\tilde{\omega}, \tilde{\Omega}, \tilde{\kappa}, \tilde{\Omega}_K\}=
\frac{r}{c_s} \{\omega, \Omega, \kappa, \Omega_K\} $ &  & $\tilde{\kappa}^2 = 2\tilde{\Omega}^2(2-q)$ & epicyclic frequency\\
 $\mu = \tilde{\omega} - m\tilde{\Omega}$ &  & $y = \tilde{\Omega}^2 - \tilde{\Omega}_K^2$ &\\
 \hline
\end{tabular}
\label{tab_notation}
\end{table*}

\subsection{Local dispersion relation}
\label{sec_disp_adiab}

We are now in a position to work towards obtaining a dispersion relation.
Note that $f_i$ and $g_i$ from equations (\ref{eq_mode1})-(\ref{eq_mode2}) are presumably different for the
different perturbed quantities. However, in the following analysis, we try to handle this issue
by eliminating entire radial derivative terms to the required order,
by combining the right equations wherever possible. First, following
\citet{1998ApJ...493..291B},  we combine the $r$ and $z$
components of the momentum balance equations (\ref{eq_radmom}) and (\ref{eq_zmom}).
Under the approximation scheme detailed in the previous section, and using
equations (\ref{eq_radeqbm}), (\ref{eq_pert_eos}) and
the notation from Table \ref{tab_notation}, these two equations can be written as respectively
\begin{equation}
 \rho_0 \biggl[ i(m\Omega - \omega)v_{1r} -2 \Omega v_{1\phi} \biggr] = -\partial_r P_1 +
 P_1\frac{\rho_0}{\gamma P_0} r(\Omega^2 - \Omega_K^2) +
 \frac{B_{0\phi}}{4\pi r}\biggl\{ i \biggl[ m + \eta \biggl(1 + \frac{l^2}{k_z^2} \biggr) \biggr]B_{1r}
 - (2 + \hat{B}_\phi)B_{1\phi}  - r\partial_r B_{1\phi}\biggr \}
 \label{appro_rmom}
\end{equation}
 and
\begin{equation}
 i\rho_0 (m \Omega - \omega) v_{1z} = -i k_z \biggl(P_1 + \frac{B_{0\phi} B_{1\phi}}{4\pi} \biggr)
 + im \frac{B_{0\phi} B_{1z}}{4\pi r} ~.
\label{appro_zmom}
\end{equation}
Next, we multiply equation (\ref{appro_rmom}) by $-i k_z$ and
differentiate equation (\ref{appro_zmom}) with respect to $r$, and add the
results. The terms involving $\partial_r( P_1 + B_{0\phi} B_{1\phi}/4\pi)$ cancel,
and then by invoking equations (\ref{eq_mode2}) and (\ref{v1z_B1z}), we obtain to the requisite order
\begin{equation}
\rho_0 \biggl[  (m\Omega - \omega) \biggl(1 + \frac{l^2}{k_z^2} \biggr)v_{1r}  + 2i \Omega v_{1\phi}\biggr] =
\frac{B_{0\phi}}{4\pi r} \biggl[ 2iB_{1\phi}   + (m +\eta) \biggl(1 + \frac{l^2}{k_z^2} \biggr) B_{1r}   \biggr]
- i P_1\frac{\rho_0}{\gamma P_0} r (\Omega^2 - \Omega_K^2) ~.
\label{eq_mcb6}
\end{equation}
Note that, with appropriate modifications, e.g.,
converting ${\bf B_1}$ to ${\bf \dot{B}_1}$, equation (\ref{eq_mcb6}) reduces to equation (3.25)
of \citet{1998ApJ...493..291B} in the absence of rotation and gravity.

For the remainder of the analysis, we can neglect the terms involving $v_{1z}$ and $B_{1z}$ in
equation (\ref{appro_zmom}) (see \S \ref{sec_approx}) such that $P_1$ can be eliminated as
\begin{equation}
P_1 = - \frac{B_{0\phi} B_{1\phi}}{4\pi} ~.
\label{eq_P1}
\end{equation}
We can also use the $r$-component of the induction equation given by equation (\ref{eq_br}),
to eliminate $B_{1r}$ in favor of $v_{1r}$ as
\begin{equation}
(m\Omega - \omega)B_{1r} = \frac{B_{0\phi}}{r} (m+\eta) v_{1r} ~.
\label{eq_B1r}
\end{equation}
We then apply equations (\ref{eq_P1}) and (\ref{eq_B1r}) to equation (\ref{eq_mcb6}).
We further define dimensionless parameters $\beta$, $\tilde{\omega}$, $\tilde{\Omega}$, $\mu$ and
${\bf b}$ as given in Table \ref{tab_notation}, in order to nondimensionalize all ensuing equations
in the present section. Thus, equation (\ref{eq_mcb6}) reduces to
%
\begin{equation}
\biggl[ \frac{\gamma \beta}{2} \mu^2 - (m + \eta)^2  \biggr] \biggl(1 + \frac{l^2}{k_z^2} \biggr) \frac{u_r}{2}
- i \mu \frac{\gamma \beta}{2} \tilde{\Omega}  u_\phi =
-i \biggl [ 1 + \frac{1}{2}(\tilde{\Omega}^2 - \tilde{\Omega}_K^2) \biggr] \mu b_\phi ~.
\label{eq_mcb11}
\end{equation}

We can now eliminate $b_\phi$ in favor of $\nabla \cdot {\bf u}$ through the continuity equation
as follows. First, we nondimensionalize equation (\ref{eq_cont}) in terms of the
new variables given in Table \ref{tab_notation} and use equation (\ref{eq_pert_eos}) to write $\rho_1$ in
terms of $P_1$. Next, we rewrite the radial equilibrium condition given by equation (\ref{eq_radeqbm})
in terms of the new variables, and using it together with
equation (\ref{eq_P1}) to replace $P_1$ in the continuity equation we obtain
%
%
%
\begin{equation}
-i \mu b_\phi = \frac{\gamma \beta}{2} r (\nabla \cdot {\bf u}) - (1 + \hat{B}_\phi) u_r +
\frac{\gamma \beta}{2} (\tilde{\Omega}^2 - \tilde{\Omega}_K^2) u_r ~.
\label{eq_mcb14}
\end{equation}
Hence, equation (\ref{eq_mcb11}) then becomes
\begin{equation}
\biggl[ \frac{\gamma \beta}{2} \mu^2 - (m + \eta)^2  \biggr] \biggl(1 + \frac{l^2}{k_z^2} \biggr) \frac{u_r}{2}
- i \mu \frac{\gamma \beta}{2} \tilde{\Omega}  u_\phi = \biggl[ 1 + \frac{1}{2}(\tilde{\Omega}^2 - \tilde{\Omega}_K^2) \biggr]
\biggl[\frac{\gamma \beta}{2} r (\nabla \cdot {\bf u}) - (1 + \hat{B}_\phi) u_r +
\frac{\gamma \beta}{2} (\tilde{\Omega}^2 - \tilde{\Omega}_K^2) u_r \biggr] ~,
\label{eq_mcb15}
\end{equation}
which reduces to equations (3.26) and (3.29) of \citet{1998ApJ...493..291B} in the non-rotating and no-gravity limit.

The $\phi$-component of the induction equation given by equation (\ref{eq_bphi})
can be simplified (to requisite order) by using equations (\ref{eq_B1r}) and (\ref{eq_mcb14}) as
%
\begin{equation}
i (m + \eta) u_\phi = \biggl( 1 + \frac{\gamma \beta}{2} \biggr) r (\nabla \cdot {\bf u}) -
\biggl[2 - \biggl(\frac{m+\eta}{\mu} \biggr) \tilde{\Omega} \frac{\partial \ln \Omega}{\partial \ln r}
- \frac{\gamma \beta}{2}  (\tilde{\Omega}^2 - \tilde{\Omega}_K^2)   \biggr] u_r ~.
\label{eq_mcb17}
\end{equation}
The above equation reduces to equation (3.28) of \citet{1998ApJ...493..291B} in
the non-rotating and no-gravity limit.

Under our approximation scheme, the $\phi$-component of the momentum balance equation given
by equation (\ref{eq_phimom}) can be written after
using equations (\ref{eq_B1r}) and (\ref{eq_mcb14}) as
%
%
\begin{equation}
i \mu^2 u_\phi = (m + \eta) r  (\nabla \cdot {\bf u}) +
\biggl[ \frac{\tilde{\kappa}^2}{2\tilde{\Omega}} \mu + (m +\eta)  (\tilde{\Omega}^2 - \tilde{\Omega}_K^2)
\biggr] u_r ~,
\label{eq_mcb19}
\end{equation}
where $\kappa^2  = 2 \Omega^2 (2 + \partial \ln \Omega/ \partial \ln r) $ is the epicyclic frequency
and $\tilde{\kappa}$ the corresponding dimensionless parameter defined in Table \ref{tab_notation}.
Equation (\ref{eq_mcb19})
reduces to equation (3.30) of \citet{1998ApJ...493..291B} in the non-rotating and no-gravity limit.

Equations (\ref{eq_mcb15}), (\ref{eq_mcb17}) and (\ref{eq_mcb19})
form a system of three homogeneous linear equations in the variables
$\{ u_r, r(\nabla \cdot {\bf u}),  i u_\phi \}$. To further simplify the notation, we
introduce the parameters $q$, $n$, $x$ and $y$, which are also defined in Table \ref{tab_notation}.
Equations (\ref{eq_mcb15}), (\ref{eq_mcb17}) and (\ref{eq_mcb19})  then become:
\begin{equation}
\biggl[ \frac{1}{2} (x \mu^2 - n^2) \biggl( 1 + \frac{l^2}{k_z^2}  \biggr)  +
\biggl(1 + \frac{y}{2} \biggr) (1 + \hat{B}_\phi - xy)    \biggr]  u_r
- x \biggl(1 + \frac{y}{2} \biggr) r(\nabla \cdot {\bf u}) - x \mu  \tilde{\Omega} (i u_\phi) = 0 ~,
\label{eq_mcb21}
\end{equation}
\begin{equation}
\biggl(2 + q \frac{n}{\mu}  \tilde{\Omega}  - xy \biggr) u_r - (1+x)r(\nabla \cdot {\bf u}) + n (i u_\phi) = 0
\label{eq_mcb22}
\end{equation}
and
\begin{equation}
 \biggl[ (2-q) \tilde{\Omega} \mu +  ny  \biggr] u_r + n r(\nabla \cdot {\bf u}) - \mu^2  (i u_\phi) = 0 ~.
 \label{eq_mcb23}
\end{equation}
We can eliminate $r(\nabla \cdot {\bf u})$ in equation (\ref{eq_mcb21})
by using equation (\ref{eq_mcb23}) to obtain
%
\begin{equation}
\biggl[ \frac{1}{2} (x \mu^2 - n^2) \biggl( 1 + \frac{l^2}{k_z^2}  \biggr)  +
\biggl(1 + \frac{y}{2} \biggr) \biggl(1 + \hat{B}_\phi  +  \frac{2-q}{n} x \tilde{\Omega} \mu \biggr)    \biggr]  u_r
- \biggl[ \tilde{\Omega}   + \frac{\mu}{n} \biggl(1 + \frac{y}{2} \biggr) \biggr] x \mu (i u_\phi) = 0 ~.
\label{eq_mcb26}
\end{equation}
We next express $u_\phi$ in terms of $u_r$ by eliminating $r (\nabla \cdot {\bf u})$
between equations (\ref{eq_mcb22}) and (\ref{eq_mcb23}). This is achieved by
multiplying equation (\ref{eq_mcb22}) by $n$ and equation (\ref{eq_mcb23}) by $(1+x)$ and
adding the results to yield
\begin{equation}
- i u_\phi =  \Biggl[ \frac{ 2 n \bigl(1 + \frac{y}{2} \bigr) + (1+x)(2-q) \tilde{\Omega} \mu
+ \frac{n^2}{\mu}q \tilde{\Omega} }{n^2 - (1+x) \mu^2}   \Biggr]  u_r ~.
\label{eq_mcb27}
\end{equation}
Finally, we substitute equation (\ref{eq_mcb27}) into equation (\ref{eq_mcb26}),
cancel the common factor of $u_r$ and after some rearrangement obtain
the (dimensionless) dispersion relation for the case including the effects of
rotation, gravity, compressibility, background toroidal and poloidal fields, magnetic curvature and
background radial gradients in an adiabatic background:
%
%
\begin{align}
\biggl[(x \mu^2 - n^2) \biggl( 1 + \frac{l^2}{k_z^2}  \biggr)  &+
(2 + y) (1 + \hat{B}_\phi) \biggr] \biggl[n^2 - (1+x) \mu^2 \biggr ] \nonumber \\
&+ 2 (2-q)x (1+x)\mu^2 \tilde{\Omega}^2
+ 4\mu x n \tilde{\Omega} (2 + y) + 2 n^2 q x \tilde{\Omega}^2
+ x \mu^2 (2 + y)^2= 0 ~.
\label{disp_mcb32c}
\end{align}

\subsection{PP05 limit of the local dispersion relation}
\label{sec_pp05_limit}

In order to compare with PP05, we need to
obtain the axisymmetric strong-$B_{0\phi}$ limit of our dispersion relation.
Hence, we put $m=0$, $n=\eta$ and
$\mu = \tilde{\omega}$ in the dispersion
relation given by equation (\ref{disp_mcb32c}), which
we further expand and divide
throughout by $x(1+x)$.
%
We also use $\tilde{\kappa}^2 = 4 \tilde{\Omega}^2 - 2q \tilde{\Omega}^2$ and
assume $x \ll 1$. On further assuming $r \partial_r P_0/(\gamma P_0) = {\cal O}(1)$ and
$(1+ \hat{B}_\phi) = {\cal O}(1)$, we
obtain from equation (\ref{eq_radeqbm})
\begin{equation}
y = \tilde{\Omega}^2 - \tilde{\Omega}_K^2 \approx \frac{(1 + \hat{B}_\phi )}{x}  \gg 1 ~.
\end{equation}
On applying all the above orderings to
equation (\ref{disp_mcb32c}), dimensionalizing it using
the definitions in Table \ref{tab_notation} and multiplying throughout by $c_s^4/r^4$,
we obtain
%
%
\begin{align}
 \biggl( 1 + \frac{l^2}{k_z^2}  \biggr) \omega^4 -
 \biggl[ k_z^2 v_{Az}^2 \biggl( 1 + \frac{l^2}{k_z^2}  \biggr)
 + \kappa^2  &+
   (1 - \hat{B}_\phi^2) \frac{v_{A \phi}^2}{r^2}\biggr]\omega^2
 - 4 \Omega (k_z v_{Az}) (1 + \hat{B}_\phi)\frac{v_{A \phi}}{r} \omega    \nonumber \\ &+
k_z^2 v_{A z}^2 \biggl[ \frac{c_s^2}{v_{A \phi}^2} \biggl \{ k_z^2 v_{A z}^2 \biggl( 1 + \frac{l^2}{k_z^2}  \biggr)
+ 2\Omega^2 \frac{\partial \ln \Omega}{\partial \ln r} \biggr \} -
(1 + \hat{B}_\phi)^2 \frac{v_{A \phi}^2}{r^2}  \biggr]  = 0 ~.
 \label{disp_eqmcb39_dim}
\end{align}
Before we compare the above dispersion relation with that obtained by PP05, we point out
that we nondimensionalize our equations differently than PP05. PP05
assumed their background equilibrium flow to be purely Keplerian, instead of considering
equation (\ref{eq_radeqbm}), which in the absence of
radial gradients becomes
\begin{equation}
\Omega^2 = \Omega_K^2 + \frac{v_{A\phi}^2} {r^2} ~,
\label{eq_radeq_nograd}
\end{equation}
i.e., PP05 ignored the magnetic tension term in the background equilibrium
flow, which becomes important when the toroidal field is
strong. However, note that they did include the effect of curvature due
to magnetic tension in their linear stability analysis (as do we).
We, on the other hand, self-consistently
nondimensionalize all the frequencies with respect to
the local $\Omega_K$, all length scales by the
local radial co-ordinate $r_0$, all wavenumbers by $1/r_0$ and all velocities by $r_0 \Omega_K$.
Then, the dimensionless equilibrium condition (\ref{eq_radeq_nograd}) becomes $\Omega^2 = 1 + v_{A\phi}^2$
(instead of $\Omega^2 = 1$, as in PP05),
the term  $2\Omega^2 \partial \ln \Omega/\partial \ln r$ becomes $-3- 2 v_{A\phi}^2$
(instead of $2\Omega^2 \partial \ln \Omega/\partial \ln r = -3$, as in PP05) and
the dimensionless epicyclic frequency becomes $\kappa^2 = 1 + 2 v_{A\phi}^2$ (instead of $\kappa^2=1$, as in PP05).
Thus, using these, the dimensionless version of equation (\ref{disp_eqmcb39_dim}),
in the limit $\hat{B}_\phi=0$, becomes
\begin{align}
   \biggl( 1 + \frac{l^2}{k_z^2}  \biggr) \omega^4 -
\biggl[ 1 + k_z^2 v_{Az}^2  \biggl( 1 + \frac{l^2}{k_z^2}  \biggr)  +   3v_{A \phi}^2 \biggr] \omega^2
 &- 4 (k_z v_{Az})  (1 + v_{A\phi}^2)^{1/2} v_{A \phi} \omega   \nonumber \\ &+
k_z^2 v_{A z}^2 \biggl[ \frac{c_s^2}{v_{A \phi}^2}  \biggl \{ k_z^2 v_{A z}^2  \biggl( 1 + \frac{l^2}{k_z^2}  \biggr)
-3 - 2 v_{A \phi}^2 \biggr \} - v_{A \phi}^2 \biggr]  = 0 ~.
 \label{disp_eqPP41_cor}
\end{align}
As a result of this difference in normalization, PP05
obtained higher growth rates for the
unstable modes at stronger field strengths (see Figure \ref{fig_Imw_all} below and compare
the growth rates with those in Figure 2 of PP05).

First, let us compare the above dispersion relation with that given by equation (25) of PP05
in the limit $l^2/k_z^2 \rightarrow 0$ (note that this limit does not violate the WKB
approximation in our case as it only implies the special case
$l^2 \ll k_z^2$ but $l^2r^2 \gg 1$ still holds true). On
dividing equation (25) of PP05 throughout by $k_z^2 v_{A\phi}^2$, neglecting the fast modes
by dropping the $\omega^6$ term and by considering the strong field limit of
$v_{A\phi}^2 \gg \{c_s^2, v_{Az}^2\}$, one arrives at a fourth degree dispersion relation
in $\omega$.
This is given by equation (41) of PP05,
which is equivalent to our equation (\ref{disp_eqPP41_cor}) above, when the former is
dimensionalized according to our aforementioned scheme and the curvature terms therein set to
$\epsilon_i = 1$, with $i=1,2,3,4$.

Next, we compare our dispersion relation given by equation (\ref{disp_eqPP41_cor}) above with
equation (25) of PP05 in the limit $l/k_z$ is finite. We note that equation (25) of PP05 has some
imaginary terms proportional to $i l/k_z^2$.
These terms, however, do not appear in our equation (\ref{disp_eqPP41_cor}), which was derived self-consistently
by retaining only the leading order WKB terms.
In this context, we refer to Appendix A and Figure 12 of PP05,
where they discussed the effect of a finite and constant $l/k_z$ on their solutions. As the
value of $l/k_z$ increased,
PP05 found a new instability having a constant growth rate independent of $k_z$, which they attributed to the
terms proportional to $i l/k_z^2$ in their dispersion relation.
Note that we do not observe any
such instability when solving with a finite $l/k_z$ in our dispersion relation given
by equation (\ref{disp_eqPP41_cor}), nor do they appear
in the global eigenvalue solutions (see Figure \ref{fig_Imw_all} below).



\section{Global eigenvalue analysis}
\label{sec_global}

\subsection{Details of the numerical set-up}
\label{sec_eig_setup}

Note that since the primary aim of this work is to compare with the local analysis in
the PP05 limit, we
restrict ourselves to axisymmetric perturbations (i.e., $\partial_\phi =0$ or $m=0$) and
ignore any radial stratification ($\partial_r \rho_0 = \partial_r B_{0\phi} = \hat{B}_\phi = 0$).
However,
both of these effects can be included in a
numerical set-up similar to that described below, which will be the focus of our next work
in this series.

For axisymmetric perturbations, the condition $\nabla \cdot {\bf B_1} =0$ is
easily imposed by invoking the vector potential ${\bf A}$ with components $(A_r, A_\phi, A_z)$,
and perturbing it such that ${\bf A} = {\bf A_0} + {\bf A_1}$. This leads to
$B_{1r} = - \partial_z A_{1\phi} $ and $B_{1z} = (1/r) \partial_r (r A_{1\phi})$, where $ A_{1\phi}$
is the azimuthal component of the perturbed vector potential.
This simplifies the system of equations (\ref{eq_cont})-(\ref{eq_bz}) by eliminating one
variable and, hence, the need for equation (\ref{eq_bz}).

Now, in order to construct the global solutions from the linearized axisymmetric
system of equations (\ref{eq_cont})-(\ref{eq_bphi}),
we consider an annular section of the accretion disc, which exhibits differential rotation
and is bounded in
radius $r \in [R_{\rm in}, R_{\rm out}]$, such that all the radial curvature terms can be included.
This region is assumed to be located far from the influence of the central
object as well as from the exteriors of the disc.
The solutions are decomposed on a basis of Chebyshev polynomials along the radial grid
and on a Fourier basis in the (local) vertical direction characterized
by the vertical wavenumber $k_z$.
We solve equations (\ref{eq_cont})-(\ref{eq_bphi}) as a linear eigenvalue problem using
Dedalus\footnote{Dedalus is available at
\url{http://dedalus-project.org}.}, an open-source pseudo-spectral code
(Burns et al., in preparation).
Recall that a linear eigenvalue problem in one dimension has the general form:
\begin{equation}
 {\cal M}(r) {\boldsymbol \xi}(r)  = \omega {\cal I} {\boldsymbol \xi}(r) ~,
 \label{eq_eigen}
\end{equation}
where $\omega = \omega_R + i \omega_I$ is the complex eigenvalue; ${\cal I}$ the identity
operator; ${\cal M}(r)$ the MHD linear differential operator and ${\boldsymbol \xi}(r)$ the eigenfunction
constituted of the perturbed variables in the system.
According to our convention for the perturbations given by equation (\ref{eq_mode1}),
the growth rate of an unstable mode is given by $\omega_I$, such that modes having
$\omega_I>0$ are unstable and those having $\omega_I \leq 0$ are stable.
The presence of a non-zero real part $\omega_R$ is an indication of the
overstability of the mode. Note that although $\omega_R=0$
implies a non-propagating mode, a non-zero $\omega_R=0$ need not
necessarily result in a
traveling mode, which in fact requires a non-zero group velocity.




Note that
Dedalus does not allow over-specification of the boundary conditions and, hence, one needs to supply exactly
as many boundary conditions as there are first-order, independent Chebyshev derivatives (which in our case
is $\partial_r$ only). For the axisymmetric problem defined above, this number is four.
We choose rigid (or hardwall) radial boundary conditions such that
\begin{equation}
v_{1r}  = 0  ~~~ {\rm at} ~~~ r = R_{\rm in}, R_{\rm out}  \nonumber
\end{equation}
\begin{equation}
\partial_r B_{1z} = 0  ~~~ {\rm at} ~~~ r = R_{\rm in}, R_{\rm out} ~,
\label{eq_bcded}
\end{equation}
where the second condition is motivated by \citet{2004ApJ...602..892K}.

We choose our fiducial case to have a resolution of $N_r=150$, where $N_r$ is the number of
radial grid points, and a wide annulus such that $R_{\rm in}=1$ and $R_{\rm out}=5$.
A few higher resolution runs, having $N_r=200$ and $256$, were also conducted as required
(see \S \ref{subsec_supra} below).
We briefly mention here how we eliminate numerically spurious eigenvalues,
which may creep in due to the truncation of the Chebyshev polynomial series at a finite $N_r$.
We essentially solve the eigenvalue
problem at two different resolutions, $N_r$ and $1.5N_r$, and retain only those modes which
are the same between the two resolutions for an assigned tolerance \citep[see, e.g.,][]{2001cfsm.book.....B}.
This ensures that the solutions are well-resolved.


We also constructed solutions for a narrow annulus
such that $R_{\rm in}=1$ and $R_{\rm out}=1.5$,
with resolutions $N_r = 64$ and $128$.
However, the results of the narrow case were only in partial accord with those of the wide case.
Recall that the {\it most} unstable modes in standard weak-field MRI are {\it all} known to be localized
close to the inner radial boundary in global analyses \citep{1994ApJ...434..206C,2015MNRAS.453.3257L}.
However, in the presence of a suprathermal toroidal field, the most
unstable modes of the system behave very differently and only {\it some} of them are localized
close to the inner radius, whereas others
span a wider radial extent (as discussed in \S \ref{sec_efuncs}).
Thus, the narrow case solutions agree with only
those wide case solutions that are localized towards the inner radial boundary.
We present here the results for the wide annulus only, with the aim of extracting the complete global picture.
In order to confirm that our fiducial annulus of $r \in [1,5]$ is indeed adequate to
capture all the global solutions,
we also conducted a few calculations using a wider annulus of $r \in [1,7]$ and $N_r = 256$. We found that
the results of the two cases were in excellent agreement.

We mention in this context that we also solved the eigenvalue problem using the
method described in Appendix A of
\citet{2016A&A...589A..87B}, which again uses a pseudo-spectral representation based on Chebyshev
polynomials in the radial direction and a Fourier basis in the vertical direction. We found these
results to be in good accord with those obtained from Dedalus, however, we present here only the latter
due to its comparatively wider scope.

\subsection{Normalization scheme}

In order to nondimensionalize our original dimensionful, linearized
equations (\ref{eq_cont})-(\ref{eq_bphi}) for the global analysis,
we use quantities at the inner radial boundary $R_{\rm in} = 1$.  We scale all frequencies by the
inner Keplerian frequency $\Omega_{\rm in}$ and
all velocities (including Alfv\'{e}n velocities) by the inner Keplerian velocity $v_{\rm in}$, which are essentially,
$\Omega_{\rm in} = v_{\rm in} = \sqrt{GM} = 1$;
all lengthscales by $R_{\rm in} = 1$; all wavenumbers by $1/R_{\rm in}$ and
all densities by the constant background density $\rho_0=1$. We also assume an adiabatic background such that
equation (\ref{eq_pert_eos}) holds true with a constant sound speed $c_s$.
We consider the exact
background equilibrium given by equation (\ref{eq_radeq_nograd}), which when nondimensionalized
according to this scheme becomes
\begin{equation}
\Omega = r^{-3/2}(1 + r v_{A\phi}^2)^{1/2} ~,
\label{dimles_radeq}
\end{equation}
where $v_{A\phi}$ now represents the dimensionless background toroidal Alfv\'{e}n velocity.
Thus, the final set of dimensionless, axisymmetric, linearized equations that we solve are

\begin{equation}
-i \omega \rho_1 + \partial_r v_{1r} + \frac{1}{r} v_{1r} + i k_z v_{1z} = 0 ~,
\label{dimles_cont}
\end{equation}
\begin{equation}
-i\omega v_{1r} - 2 r^{-3/2} (1 + r v_{A\phi}^2 )^{1/2} v_{1 \phi} + c_s^2 \partial_r \rho_1
- v_{A\phi}^2 \frac{\rho_1}{r} + v_{A\phi} \partial_r v_{A 1\phi}
+ v_{A z} \biggl( \partial_r^2 a_{1 \phi} + \frac{1}{r} \partial_r a_{1 \phi} -
\frac{a_{1 \phi}}{r^2}  -  k_z^2 a_{1\phi} \biggr)
+  \frac{2}{r} v_{A\phi} v_{A 1\phi} = 0 ~,
\label{dimles_vr}
\end{equation}
\begin{equation}
-i \omega v_{1\phi} + \frac{1}{2} r^{-3/2} (1 + r v_{A\phi}^2 )^{1/2} v_{1r}
+ \frac{1}{2}  r^{-1/2} v_{A\phi}^2 (1 + r v_{A\phi}^2 )^{-1/2} v_{1r} - i k_z  v_{Az} v_{A1\phi}
+ \frac{1}{r} i k_z  v_{A\phi} a_{1 \phi} = 0  ~,
\label{dimles_vphi}
\end{equation}
\begin{equation}
-i \omega v_{1z} + i k_z c_s^2 \rho_1 + i k_z v_{A\phi} v_{A1 \phi} = 0 ~,
\label{dimles_vz}
\end{equation}
\begin{equation}
\omega k_z a_{1\phi} + i k_z v_{Az} v_{1r} = 0  ~,
\label{dimles_br}
\end{equation}
\begin{equation}
-i \omega v_{A1 \phi}  -i k_z v_{Az} v_{\phi} - \frac{3}{2} r^{-3/2} (1 + r v_{A\phi}^2 )^{1/2} i k_z a_{1 \phi}
+ \frac{1}{2} r^{-1/2} v_{A\phi}^2 (1 + r v_{A\phi}^2 )^{-1/2} i k_z a_{1 \phi}
+ v_{A\phi} (\partial_r v_{1r} + i k_z v_{1z})   = 0  ~,
\label{dimles_bphi}
\end{equation}
where,
\begin{equation}
v_{A1 \phi} = \frac{B_{1\phi}}{\sqrt{(4 \pi \rho_0)}} \biggl(\frac{1}{v_{\rm in}} \biggr) ~,
\end{equation}
\begin{equation}
a_{1 \phi} = \frac{A_{1\phi}}{\sqrt{(4 \pi \rho_0)}} \biggl(\frac{1}{v_{\rm in} R_{\rm in}} \biggr)
\end{equation}
and $v_{Az}$ is the dimensionless vertical Alfv\'{e}n velocity.

The eigenfunctions ${\boldsymbol \xi}(r)$, from equation (\ref{eq_eigen}), for the above problem are then given by
$\{\rho_1, v_{1r}, v_{1\phi}, v_{1z}, v_{A1\phi}, a_{1\phi}\}$.
Note that all the variables appearing henceforth in this work (i.e., in both the global eigenvalue
solutions and the PLUTO simulations)
are nondimensionalized according to scheme explained in this section.


\section{Global Solutions}
\label{sec_global_solns}

Here we present the solutions of the global eigenvalue problem
defined by equations (\ref{dimles_cont})-(\ref{dimles_bphi}).
Note that {\it both} the global and local solutions
include the effects of magnetic curvature and compressibility, while ignoring
all background gradients except for the radially varying angular velocity.
Following PP05, we fix the dimensionless sound speed at $c_s = 0.05$ and
the dimensionless vertical Alfv\'{e}n velocity at $v_{Az} = 0.01$, and compute solutions
for different values of (constant) dimensionless toroidal Alfv\'{e}n velocity $v_{A\phi}$,
focusing on the suprathermal cases such that $c_s < v_{A\phi} < 1$ (we discuss the meaning 
of the upper limit on $v_{A\phi}$ in \S \ref{subsec_supra}).

Note that a reduction in the growth rate of MRI
is expected to occur with the increase of the background toroidal field strength,
as has been shown in weak-field studies even in the absence of magnetic curvature \citep{1994ApJ...421..163B}.
What PP05 were interested in is any other change in behavior due to
the inclusion of curvature, albeit in a local framework, whereas we are interested to see if their findings
hold true in a global framework.

\subsection{PP05 Stability criteria}
\label{sec_stabcrit}

Before laying out our results, we briefly recall the
instability criteria obtained by PP05 from their approximate local analysis, as a guideline for
our comparison study.
Equations (47) and (48) of PP05 give the critical wavenumbers bounding the different suprathermal
instabilities shown in Figure 5 of PP05 such that
\begin{equation}
(k_{c1} v_{Az})^2 = 3
\label{PP_kc1}
\end{equation}
and
\begin{equation}
(k_{c2} v_{Az})^2 = \frac{v_{A\phi}^4}{c_s^2} - 1 ~,
\label{PP_kc2}
\end{equation}
recalling again that PP05 considered a purely Keplerian background
equilibrium flow
with a dimensionless epicyclic frequency $\kappa =1$
to obtain these relations.
Note that the criterion given by equation (\ref{PP_kc1}) is
the classic Balbus-Hawley criterion
for MRI to operate in a Keplerian flow \citep{1991ApJ...376..214B},
and for $v_{Az}=0.01$, this yields $k_{c1} \approx 173$.
The existence of these critical
wavenumbers thus sets the stabilization scale for the instabilities in the suprathermal regime.
We note here that the various critical wavenumbers derived by PP05 are only approximate
and their exact values would depend on the full global solution of the linearized MHD equations.
Additionally, the critical wavenumber that limits MRI in the global analysis would no longer remain
a constant but would depend on
the background toroidal field. Hence, from now onwards, we refer to the maximum allowed
vertical wavenumber for MRI to occur in the global analysis as $k_{\rm MRI}$ for simplicity
(note that in the approximate local analysis of PP05, $k_{\rm MRI} \equiv k_{c1}$).

According to PP05, setting the right hand side of equation (\ref{PP_kc2}) above to zero
yields the critical value for the suprathermal Alfv\'{e}n speed above which MRI
{\it starts} to get stabilized for the smallest wavenumbers, i.e., $v_{A \phi}^{PP1} = \sqrt{c_s}$. Consequently,
setting the right hand sides of equations (\ref{PP_kc1}) and (\ref{PP_kc2}) equal to each other yields
the critical value at which MRI is {\it completely} stabilized for all wavenumbers, namely,
$v_{A \phi}^{PP2} = \sqrt{2 c_s}$. For $c_s = 0.05$,
these limits become $v_{A \phi}^{PP1} \approx 0.22$ and
$v_{ A\phi}^{PP2} \approx 0.32$.

Keeping this general picture in mind, we will now proceed
to verify to what degree our global solutions agree with the local predictions.

\subsection{Numerical eigenvalue analysis}

\subsubsection{Suprathermal instabilities}
\label{subsec_supra}

\begin{figure*}
\centering
    \includegraphics[width=\textwidth]{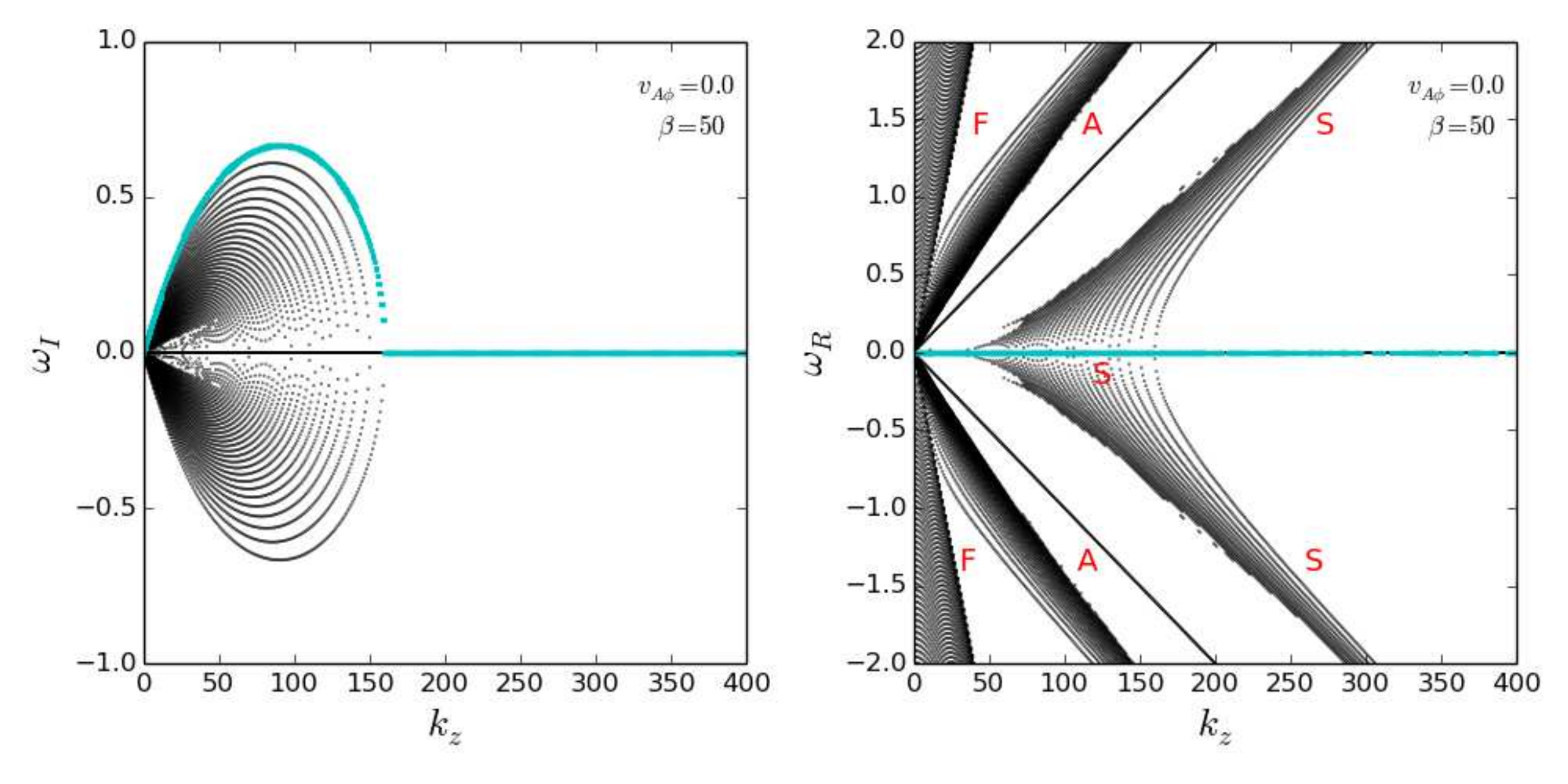} 
       \caption{Global eigenvalue solutions for standard MRI (i.e., $v_{A\phi}=0$).
       {\it Left panel}: The imaginary part $\omega_I$ of the modal frequency, or, the growth rate of all the
       unstable modes, as a function of $k_z$. {\it Right panel}:
       The real part $\omega_R$ of the modal frequency as a function of $k_z$.
       The cyan line in
       both panels demarcates the most unstable modes for this case. The letters F, S and A
       denote the fast, slow and Alfv\'{e}n modes respectively.
       The background accretion flow has $c_s=0.05$ and $v_{Az} = 0.01$,
       and the global problem is solved on a radial grid $r \in [1,5]$ with resolution $N_r=150$.}
         \label{fig_mri_growth}
\end{figure*}

We begin by showing the global eigenvalue solutions for the standard weak-field MRI case
($v_{A\phi}=0$) as a reference for the suprathermal cases ($v_{A\phi} > c_s$) to follow.
In order to estimate the associated plasma-$\beta$ of our cases, we recall that
$\beta = P/P_B = 8\pi P/(B_\phi^2+B_z^2)$. If we consider $P=\rho c_s^2$, then we can write
$\beta = 2 c_s^2 / (v_{A\phi}^2 + v_{Az}^2)$. Thus, for $v_{Az}=0.01$ and $c_s=0.05$,
the case with $v_{A\phi}=0$ has $\beta=50 \gg 1$, while the cases with $v_{A\phi}=0.1,0.25,0.3,0.4$,
have $\beta = 0.5, 0.08,0.06,0.03 < 1$ respectively.

In Figure \ref{fig_mri_growth}, we plot the imaginary part $\omega_I$ of the
eigenvalues as a function of the vertical wavenumber $k_z$, indicating
all the unstable modes for $v_{A\phi}=0$ (left panel). We also plot the corresponding
real part $\omega_R$ of all the modes (right panel), which is an indicator of the
overstability of a mode.
Note that a global problem exhibits a large family of modes as seen in Figure
\ref{fig_mri_growth}. Thus, each value of $k_z$ corresponds to multiple modes and the mode with
the {\it maximum} growth
rate is the {\it most} unstable mode at that $k_z$. The locus connecting the maximum growth rate
modes at different $k_z$
demarcates the set of {\it most} unstable modes of the system, which is indicated by
the cyan line in
the left panel of Figure \ref{fig_mri_growth}.
The cyan line in the right
panel of Figure \ref{fig_mri_growth} denotes the $\omega_R$
corresponding to the {\it most} unstable modes.

We see from Figure \ref{fig_mri_growth} that the maximum possible growth rate for standard MRI
in the global analysis
is about $0.67$, which is less than but comparable to the local prediction of $0.75$ \citep{1998RvMP...70....1B},
as also observed in previous global studies \citep[see e.g.,][]{1994ApJ...434..206C,2015MNRAS.453.3257L}.
The corresponding phase velocities of the most unstable modes are exactly zero,
which also matches the local theory, as MRI is known to be a purely unstable
mode with a non-propagating character.
We note from Figure \ref{fig_mri_growth} that $k_{\rm MRI} \approx 160$, i.e., less than the local
prediction of $\sim 173$ (see \S \ref{sec_stabcrit}).
In the right panel of Figure \ref{fig_mri_growth}, we also demarcate the bands of
fast magnetosonic, Alfv\'{e}n and slow magnetosonic modes, indicated
by the red letters F, A and S respectively. This is to better understand the nature of the {\it new}
suprathermal instabilities --- just
like in the case of standard MRI, which is known to be a destabilized slow mode.
In this context we mention that for standard weak-field MRI,
the global problem has a dispersion relation
that happens to be identical in form to that in the local problem \citep{2015MNRAS.453.3257L}.
This explains the excellent agreement between the local and the global
solutions in this case.

\begin{figure*}
\centering
 \includegraphics[width=0.9\textwidth]{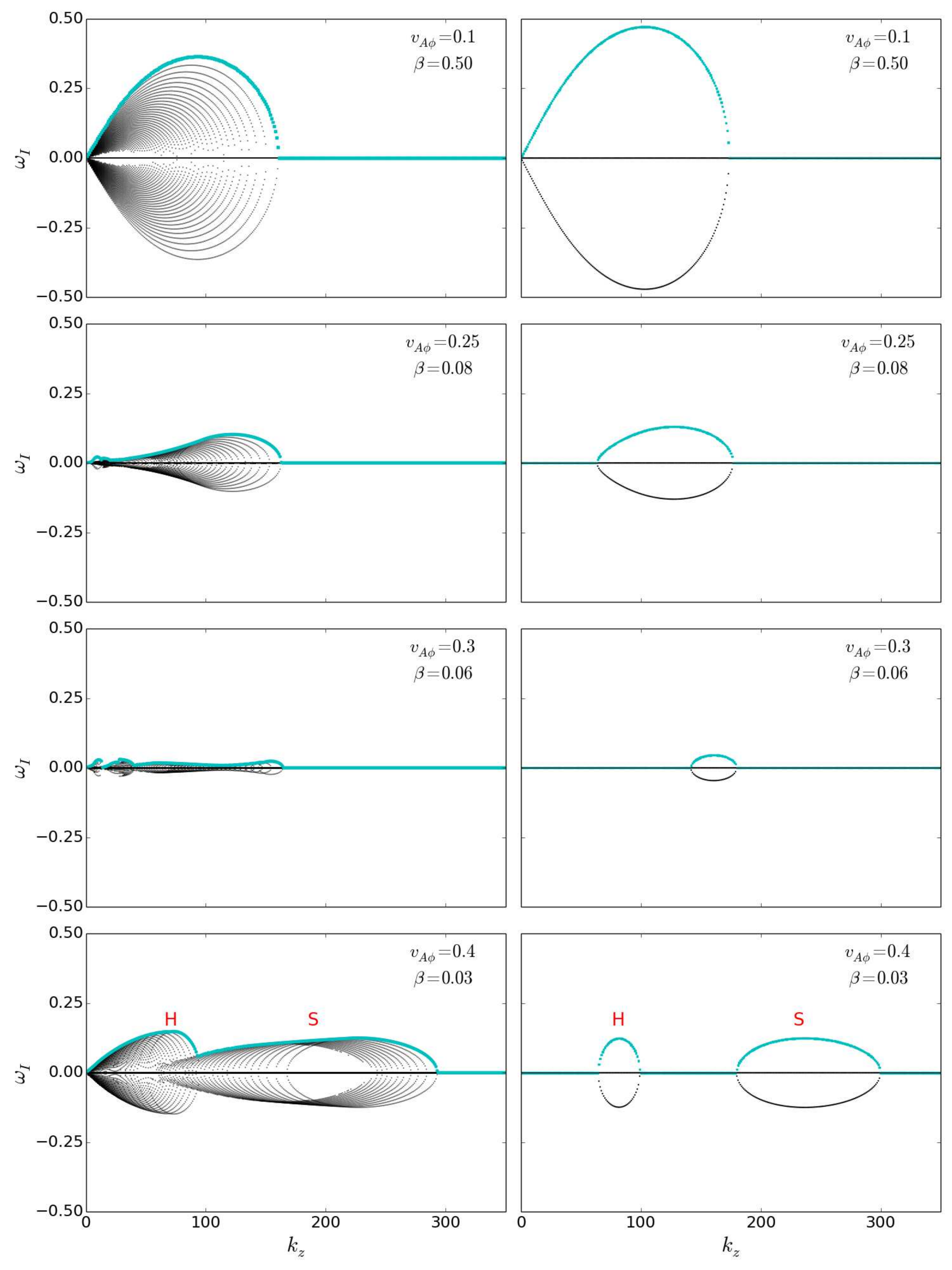}
       \caption{The growth rate $\omega_I$ of the unstable modes as a function of $k_z$ for
       different suprathermal $v_{A\phi}$ and $\beta<1$. {\it Left panel}: Global eigenvalue solutions.
       {\it Right panel}: Solutions of the local dispersion relation given by
       equation (\ref{disp_eqPP41_cor}) with $l=10$.
       The cyan line in all the panels demarcates the most unstable modes for the corresponding $v_{A\phi}$.
       The letters H and S denote hybrid and slow modes respectively.
       The background accretion flow has $c_s=0.05$ and $v_{Az} = 0.01$,
       and the global problem is solved on a radial grid $r \in [1,5]$ with resolution $N_r=150$.}
         \label{fig_Imw_all}
\end{figure*}

\begin{figure*}
\centering
 \includegraphics[width=0.9\textwidth]{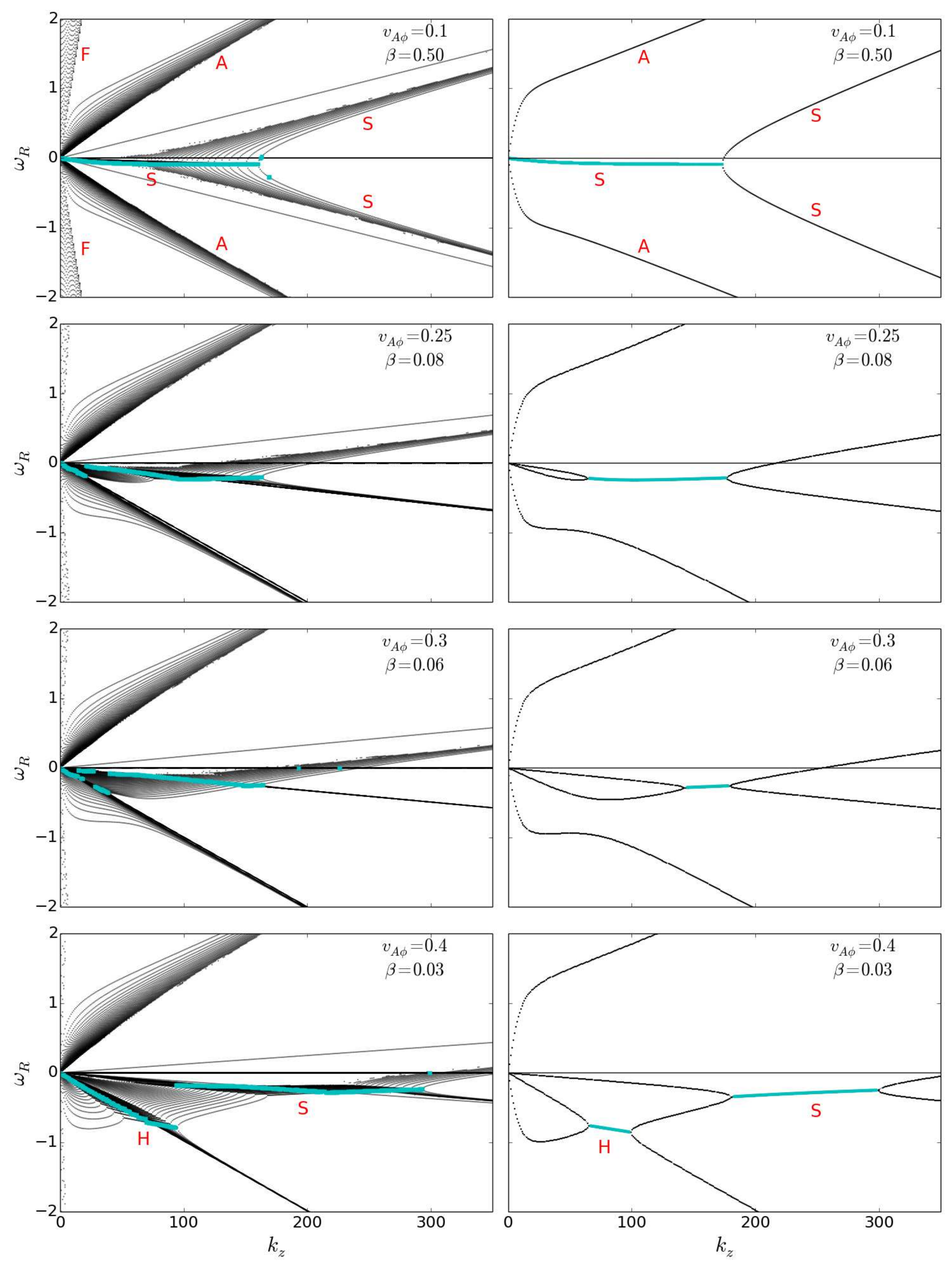}
       \caption{The real part $\omega_R$ of the modal frequency as a function of $k_z$ for
       different suprathermal $v_{A\phi}$ and $\beta<1$. The left and right panels represent the same
       cases as the left and right panels of Figure \ref{fig_Imw_all}.
       The letters F, S, A and H denote the fast, slow, Alfv\'{e}n and hybrid modes respectively.
       The cyan line in all the panels denotes the $\omega_R$ corresponding to
       the most unstable modes for the corresponding $v_{A\phi}$.
       The background accretion flow has $c_s=0.05$ and $v_{Az} = 0.01$,
       and the global problem is solved on a radial grid $r \in [1,5]$ with resolution $N_r=150$.}
         \label{fig_Rew_all}
\end{figure*}

One of the goals of the current work is to study
the evolution of the {\it most} unstable modes of the system as the
background toroidal field becomes suprathermal, the results of which are
summarized by Figures \ref{fig_Imw_all}
and \ref{fig_Rew_all}.
Figure \ref{fig_Imw_all} compares the growth rate $\omega_I$ as a
function of $k_z$ obtained from the {\it global} eigenvalue solutions (left panels)
of the axisymmetric equations (\ref{dimles_cont})-(\ref{dimles_bphi}) with that
obtained from the solutions of the {\it local} dispersion
relation (right panels) given by equation (\ref{disp_eqPP41_cor}) for
a constant radial wavenumber $l=10$,
for different suprathermal $v_{A \phi}$.
This choice of $l$ satisfies the WKB constraint (see \ref{sec_approx}) and, also,
ensures that the value of $l^2/k_z^2$ remains small for
a large range of $k_z$, which is essentially the PP05 limit we want to compare
our global results with.
Figure \ref{fig_Rew_all} shows the real part of the eigenvalues $\omega_R$
as a function of $k_z$ for the same cases as in Figure \ref{fig_Imw_all}.
Note that we mainly compare the {\it most}
unstable modes (denoted by cyan lines) between the left and
right panels (or PP05 case) of these two figures.
We list the main results of this section below:

\begin{enumerate}

\item At a glance, we observe that the {\it maximum} growth rates
for different $v_{A\phi}$ appear to be similar between the left and right panels
of Figure \ref{fig_Imw_all},
with the latter slightly overestimating the growth rates in general
(except for $v_{A\phi} = 0.4$, as we note below).
The $\omega_R$ of the most unstable modes (cyan line) also looks similar
between the left and right panels of Figure \ref{fig_Rew_all}.
Additionally, we observe that the band of
fast modes indeed lie decoupled from the rest of the modes in the global case,
thus justifying our assumption of neglecting them in \S \ref{sec_approx}.
However, as $v_{A\phi}$ increases, the global and local solutions start differing in the
vertical wavenumber occupancies of the most unstable modes.
Note from Figure \ref{fig_Imw_all} that we do not observe any modes growing on
arbitrarily short scales as PP05 find for a finite $l/k_z$, either in our local or
global solutions, as discussed in \S \ref{sec_pp05_limit}.

\item For $v_{A\phi} = 0.25$, which is slightly greater than the first PP05 limit
($v_{A \phi}^{PP1} \sim 0.22$), the most unstable modes in the global case exhibit a
low growth rate tail
extending all the way to $k_z=0$, unlike in the corresponding PP05 case,
which shows complete MRI stabilization at low $k_z$. The $\omega_R$ of the
most unstable modes in the global case also extends all the way to $k_z=0$.
In order to ensure that this result is not an artifact of inadequate
resolution, we verified it using two higher resolutions, $N_r=200$ and $256$.

\item When $v_{A\phi} = 0.3$, i.e., close to the second PP05 limit
($v_{A \phi}^{PP2} \sim 0.32$), we observe
a drastic suppression of the MRI growth rate in the global case,
to only a few percent of the maximum
growth rate for standard MRI (compare with the left panel of Figure \ref{fig_mri_growth}),
as predicted by PP05.
Thus, there indeed seems to be a threshold $v_{A\phi}$ for MRI to operate.
However, even at this stage, the instability extends all the way to
the smallest vertical wavenumbers (as displayed by the corresponding cyan lines in the global
case of Figures \ref{fig_Imw_all} and \ref{fig_Rew_all}). This is
contrary to the PP05 case
where the MRI is suppressed across almost the entire range of allowed wavenumbers.
The global result was further verified using two higher resolutions of
$N_r=200$ and $256$.

\item When the toroidal field strength increases further,
there seems to be the appearance of two new instabilities for $v_{A\phi} =0.4$,
in the local as well as the global analysis.
We note here that although PP05 mention the origin
of the two new instabilities ``as the result of coupling between
the modes that become the Alfv\'{e}n and the slow modes in the limit of no
rotation", this does not provide a very clear insight into the
exact nature of these instabilities. We attempt to resolve this issue here.

Recalling from the right panels of Figure  \ref{fig_Imw_all}, the
instability in the region $k_{c1} < k_z \leq k_{c2}$
was termed instability II by PP05,
which they claimed to be a
generalization of the axisymmetric toroidal buoyancy (ATB) mode proposed by
\citet[][]{2000ApJ...540..372K}.
The instability that emerges in a {\it limited} range
$0< k_z < k_{c1}$, was termed termed instability III by PP05 (see \S \ref{sec_stabcrit}
for the definitions of the various critical wavenumbers).
Now, if we track the slow modes as $v_{A\phi}$
increases down the right panels of Figure \ref{fig_Rew_all}, we note from the bottommost
panel that instability II of PP05 seems to arise due to a destabilization of the same
slow modes that give rise to MRI. However, since this instability appears only for suprathermal magnetic
fields $v_{A\phi} >0.3$, we refer to it with the more physically motivated
name of ``suprathermal slow mode instability (SSMI)".
This slow mode is denoted by the letter S in the bottommost right panels of Figures \ref{fig_Imw_all}
and \ref{fig_Rew_all}.
Similarly, if we track the Alfv\'{e}n modes along with the slow modes,
we notice a merging of these two modes in the bottommost right panel of Figure \ref{fig_Rew_all},
in a limited wavenumber range $0<k_z<k_{c1}$.
This merging indicates the loss
of the individual identities of the slow and Alfv\'{e}n modes and the
emergence of a hybrid unstable mode. This mode is labeled with an H
in the bottommost right panels of Figure \ref{fig_Imw_all}
and \ref{fig_Rew_all}.
We refer to it as the
``suprathermal hybrid mode instability (SHMI)" instead of instability III,
in order to convey its physical origin.
If we track the band of
slow and Alfv\'{e}n modes down the left panels of
Figure \ref{fig_Rew_all}, we see that the same deduction is applicable
in the global case as well (for the most unstable modes).
In this context, we mention that \citet{2001ApJ...553..987B}
also studied the interplay of various Alfv\'{e}nic,
slow and fast mode instabilities, however, only in weakly magnetized,
radiation pressure dominated accretion discs.

\item We observe, however, a few differences between the local and global solutions
for $v_{A\phi} =0.4$.

Both SSMI and SSMI have roughly the same maximum growth rate for $v_{A\phi} = 0.4$
in the local case. However, in the global case, the maximum growth rate of
SHMI is slightly higher than that of SSMI.
Also, in the global
case, SHMI extends all the way to $k_z=0$ unlike in the PP05 case.
This seems to point to the fact that SHMI is the mode that replaces MRI beyond the second PP05 limit.
Next, there seems to be an overlap between
the {\it most} unstable modes (cyan line) of SHMI and SSMI in the $\omega_I-k_z$ plane.
The growth rate of the most unstable modes of SHMI drops at $k_z \sim 93$ but rises again, and smoothly
gives way to the most unstable modes of
SSMI well before going to zero. This is in direct contrast with the corresponding PP05 case,
where SHMI gets completely stabilized at $k_z \sim 100$ before SSMI appears at a distinctly higher $k_z$
(see bottommost right panel of Figure \ref{fig_Imw_all}).
However, we observe a clear discontinuity in the $\omega_R-k_z$ plane at $k_z \sim 93$
for the {\it most} unstable modes (cyan lines) in the global case of Figure \ref{fig_Rew_all}.
This establishes
that SHMI and SSMI are not the same but distinct instabilities,
having different origins, different phase velocities and occupying different wavenumber regimes.


\item We summarize here the different critical wavenumbers that
emerge from the
global analysis in the suprathermal regime (as seen in the left panels of
Figures \ref{fig_Imw_all} and \ref{fig_Rew_all}).
Note that for all $v_{A\phi} \lesssim 0.3$,
the maximum allowed vertical wavenumber for the instability is still given
by $k_{\rm MRI}$ (in accordance with the Balbus-Hawley instability criterion)
and MRI, although suppressed, occupies the {\it entire} range $0 \leq k_z \leq k_{\rm MRI}$.

With a further increase in $v_{A\phi}$, two new critical wavenumbers emerge.
The larger of the two critical wavenumbers is the maximum allowed wavenumber
for SSMI, which matches reasonably well with that predicted
by PP05, namely, $k_{c2}$ given by equation (\ref{PP_kc2}).
Thus, SSMI does not keep growing indefinitely
with increasing $k_z$ in the global case but stabilizes at a well defined short
scale.
We term the smaller critical wavenumber $k_{\rm SI}$ (where SI stands for suprathermal instability),
which denotes the point of discontinuity in the $\omega_R-k_z$ plane for the most unstable
modes when $v_{A\phi}=0.4$.
This demarcates {\it both} the maximum allowed wavenumber for SHMI
as well as the minimum onset wavenumber for SSMI.
(Note that $k_{\rm SI}$ is a characteristic of only the {\it most} unstable SHMI and
SSMI modes, unlike $k_{c2}$, which denotes the maximum limit for {\it all} SSMI modes.)
Interestingly,
unlike $k_{c2}$, the critical wavenumber $k_{\rm SI}$ does not have a local analog.
Thus, the most unstable modes of SHMI lie in the
wavenumber range $0 \leq k_z \leq k_{\rm SI}$, whereas those of SSMI occupy the wavenumber range
$k_{\rm SI} \leq k_z \leq k_{c2}$ in the global case.
We mention here that
with increasing $v_{A\phi}$, the maximum growth rates of both SHMI and SSMI increase.
These two instabilities also expand their
respective wavenumber
occupancies, such that both $k_{\rm SI}$ and $k_{c2}$ increase with an increase in $v_{A\phi}$.
The fact that in the global case, SHMI and SSMI coexist for the same $v_{A\phi}$, overlap
in the $\omega_I-k_z$ plane and span a large wavenumber range between them,
might have important implications for energy exchange
between these modes in the non-linear regime.

\item We briefly discuss in this context how the large field instability
or, LFI, of \citet{1995ApJ...453..697C} compares with the new suprathermal instabilities
obtained in this work.

The LFI is a single instability that appears after MRI stabilization
in an {\it incompressible} flow having a shear parameter $q \neq 2$,
for $v_{A\phi}$ {\it greater} than the rotational speed.
The SHMI and SSMI on the other hand are two distinct instabilities, which
appear and coexist after MRI stabilization
in a {\it compressible} flow for suprathermal $v_{A\phi}$
{\it less} than the rotational speed.
The maximum growth rate and vertical wavenumber occupancy of LFI increase with $v_{A\phi}$
(see Figure 5 of \citealt{1995ApJ...453..697C}), which is similar to the behavior of
SHMI and SSMI mentioned above. Both LFI and SHMI (in the global case) have
growth rates extending all the way to $k_z=0$ (in contrast with the local prediction).
Thus, SHMI and SSMI together can be thought of as the compressible counterpart of LFI.

\item  An important point to note here is that although our current analysis is
cylindrical and vertically unstratified, in a realistic disc $k_z$ cannot be arbitrarily
small as it is limited by the disc thickness.
Following the convention in \citet{1998RvMP...70....1B}, we assume the vertical
scale height of the disc to be $H = \sqrt{2} c_s$ (in our dimensionless units). Moreover,
at least one wavelength of the unstable mode must fit within a
disk thickness of $2H$.
Thus, instability requires $\lambda = 2 \pi/ k_z < 2H$ or
$k_z H > \pi$. On using $c_s=0.05$, we get a lower limit on $k_z$ such that $k_z \gtrsim 45$
is the physically allowed regions for instability.

We now interpret what it means when the above restriction is applied to the
global solutions in Figures \ref{fig_Imw_all} and \ref{fig_Rew_all}. Even when
the low-$k_z$ zone is discarded, the suppression of MRI with increasing suprathermal
$v_{A\phi}$ is still relevant as it occurs at all $k_z$, and this effect should be captured in
global, vertically stratified simulations. However, when $v_{A\phi}$
is increased beyond MRI stabilization, a significant portion of the low-$k_z$ region occupied
by SHMI becomes unphysical (as seen from the bottommost left panel of Figure \ref{fig_Imw_all}),
while SSMI remains largely unaffected.
Nevertheless, SHMI can operate in the range $45 < k_z \lesssim 93$ (for $v_{A\phi} =0.4$)
and, hence, might still be a physical instability. Moreover, with a further increase in
$v_{A\phi}$, SHMI expands its wavenumber occupancy and, hence, can occupy larger
physically allowed regions.

\item We find that the
most unstable modes of {\it all}
the suprathermal instabilities (i.e., for all $v_{A\phi} \geq 0.1$) are overstable, having
non-zero $\omega_R$ in both the local and global cases (unlike in standard MRI), as seen
in Figure \ref{fig_Rew_all}.
Depending on the relative magnitudes of $\omega_I$ and $\omega_R$, the unstable modes could be
predominantly amplified, if $|\omega_I| > |\omega_R|$ or predominantly
oscillatory, if $|\omega_I| < |\omega_R|$.
The group velocity of the most unstable modes in the global case can be estimated
numerically from the left panels of Figure \ref{fig_Rew_all},
such that $v_{gr} = \partial \omega_R/\partial k_z$.
A non-zero $v_{gr}$ would imply that these unstable modes travel vertically across the disc.
The distance traveled by an unstable mode in unit growth time is given by
$\Delta z = |v_{gr}|/|\omega_I|$ (note that both $v_{gr}$ and $\Delta z$ are functions of $k_z$).
However, in order to estimate how large $\Delta z$ is,
we have to compare it with $H$ (for $c_s=0.05$, $H \sim 0.07$). For example, for
$v_{A\phi}=0.1$, we obtain $0 \leq |v_{gr}| \lesssim 0.002$
and $0 \leq \Delta z \lesssim 0.02$ for the unstable MRI mode. For $v_{A\phi}=0.4$, we obtain
$0.005 \lesssim |v_{gr}| \lesssim 0.02$
and $0.02 \lesssim \Delta z \lesssim 0.05$ for SHMI, and
$0\leq |v_{gr}| \lesssim 0.001$
and $0 \leq \Delta z \lesssim 0.01$ for SSMI. Since $\Delta z < H$ in all cases,
this implies that
these modes probably do not propagate far enough, in unit
growth time, to invalidate a local analysis along the vertical direction.

\item We take a moment here to briefly discuss the meaning of the upper limit on $v_{A\phi}$
maintained in this work.
The non-Keplerian background angular velocity in our work has a rotational component that decreases with
radius, as well as a magnetic component that is constant.
The limit $v_{A\phi} < 1$ means, in our units, that the toroidal Alfv\'{e}n velocities are chosen
such that they are always less than
the Keplerian velocity at the inner radius $R_{\rm in}=1$,
i.e.,  $v_{A\phi} < v_{\rm in}=1/\sqrt{R_{\rm in}}$. This ensures that
the {\it inner} disc is always rotationally supported (i.e., the rotational
component is larger than the magnetic component in the background flow).
A more stringent upper limit, to ensure that the {\it entire} disc is rotationally
supported, is given by
$v_{A\phi} < 1/\sqrt{R_{\rm out}}$, which for
our fiducial case with $R_{\rm out}=5$ implies that $v_{A\phi} < 0.45$ should be
satisfied. If $v_{A\phi} > 1/\sqrt{R_{\rm out}}$,
then it means the outer portion of the disc is magnetically supported,
which might affect the unstable modes of the system differently. 

Since our aim in this work is to compare directly with the results of PP05, we fixed $c_s=0.05$ and
$v_{Az}=0.01$ following them, but restricted our suprathermal results to $v_{A\phi}=0.4$
to ensure a fully rotationally supported disc. However,
one can repeat the analysis for a smaller $c_s$ (and a suitably smaller $v_{Az}$ to ensure it is
subthermal), in order to explore the effect of a wider
range of suprathermal $v_{A\phi}<1$, without treading the uncertain regime
of magnetically supported discs. Hence, we
also carried out the global analysis using $c_s=0.01$ and $v_{Az}=0.005$, the results 
of which are qualitatively
similar to those in Figures \ref{fig_Imw_all} and \ref{fig_Rew_all}. The critical fields for
MRI suppression are lower and in accordance with equations (\ref{PP_kc1}) and (\ref{PP_kc2}), such that
$v_{A \phi}^{PP1} \sim 0.1$ and $v_{A \phi}^{PP2} \sim 0.14$, and the vertical wavenumbers span a wider
range due to a smaller $v_{Az}$.  

We encountered an interesting difference between the local and global solutions
when performing the analysis with a different $c_s$. In the {\it local} case, $v_{A \phi}^{PP2}$ demarcates
both the complete stabilization of MRI as well as the onset of SSMI, however, SHMI appears at a slightly higher
$v_{A\phi}$, let us call it  $v_{A \phi}^{PP3}$
(which has no analytical criteria and needs to be determined numerically). As $c_s$ decreases, the
separation between $v_{A \phi}^{PP2}$ and $v_{A \phi}^{PP3}$ keeps increasing. In the
{\it global} case, on the contrary, SHMI and SSMI always appear together, right after the suppression of
MRI at $\sim v_{A \phi}^{PP2}$.

\end{enumerate}

\begin{figure*}
\centering
 \includegraphics[width=0.84\textwidth]{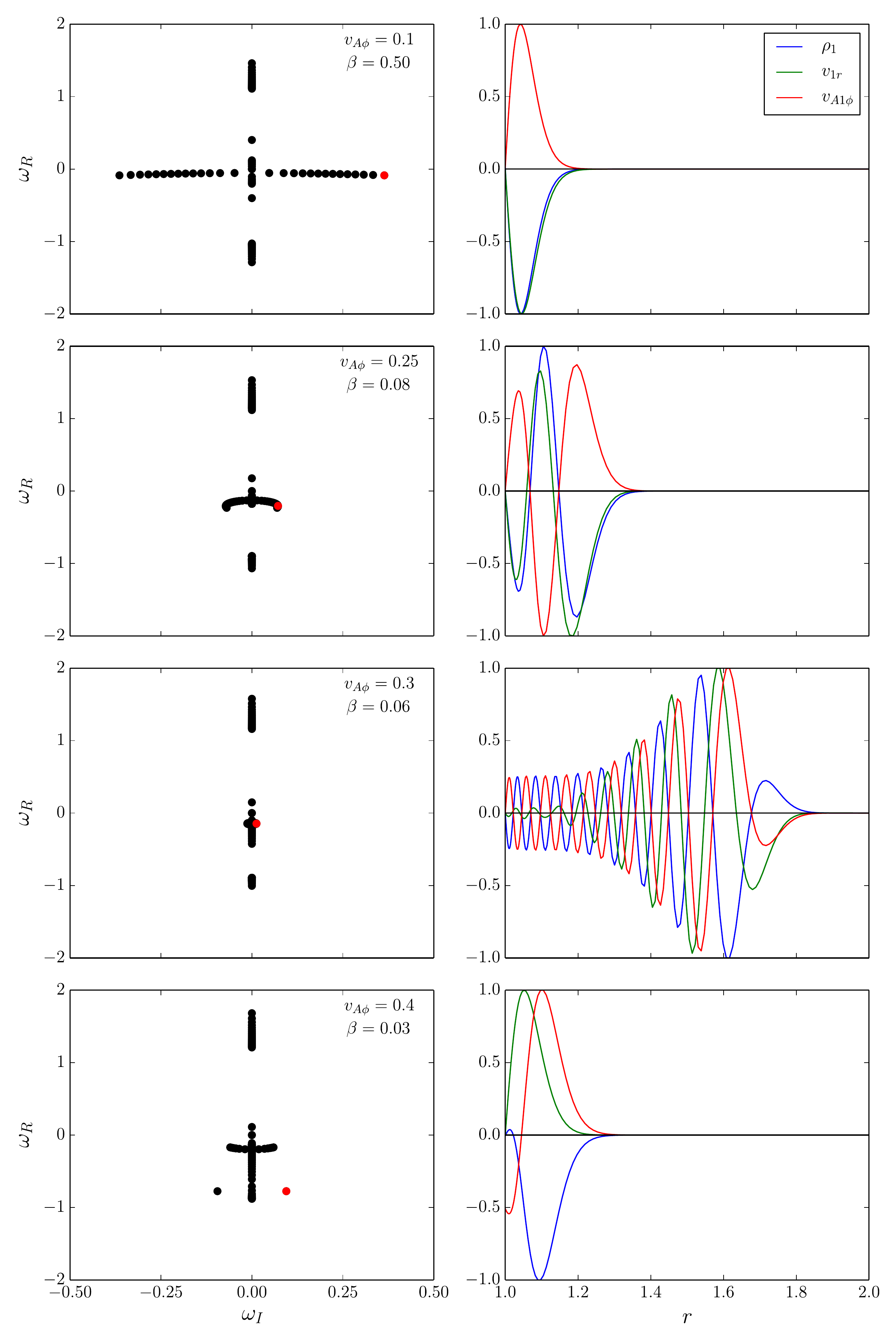}
       \caption{Global properties for different suprathermal cases computed at $k_z=90$.
       {\it Left Panel}: Eigenspectrum of all the modes at $k_z=90$, with the red dot denoting the most
       unstable mode. {\it Right panel}: Normalized eigenfunctions for the most unstable mode
       at $k_z=90$ as a function of radius $r$. The background accretion flow has $c_s=0.05$ and $v_{Az} = 0.01$,
       and the global problem is solved on
       a radial grid $r \in [1,5]$. The grid resolution is $N_r=150$ for all the left panels, and
       for the right panels it is $N_r=256$ for $v_{A\phi}=0.1,0.25,0.4$ and $N_r=512$ for $v_{A\phi}=0.3$.
       Note that the radial axis in the right panel has been zoomed close to the inner boundary.}
         \label{fig_efunckz90}
\end{figure*}

\subsubsection{Global eigenfunctions and eigenspectra}
\label{sec_efuncs}

A very important benefit of the global problem
over a local analysis is that it
allows us to study the global eigenfunctions of the problem.
In this section, we study the radial eigenfunctions
for a {\it fixed} $k_z$ and {\it different} $v_{A\phi}$.
We also look at the complete eigenspectra of
all the modes (both stable and unstable), again
for a fixed $k_z$ and different $v_{A\phi}$.



\begin{figure*}
\centering
       \includegraphics[width = 0.84\textwidth]{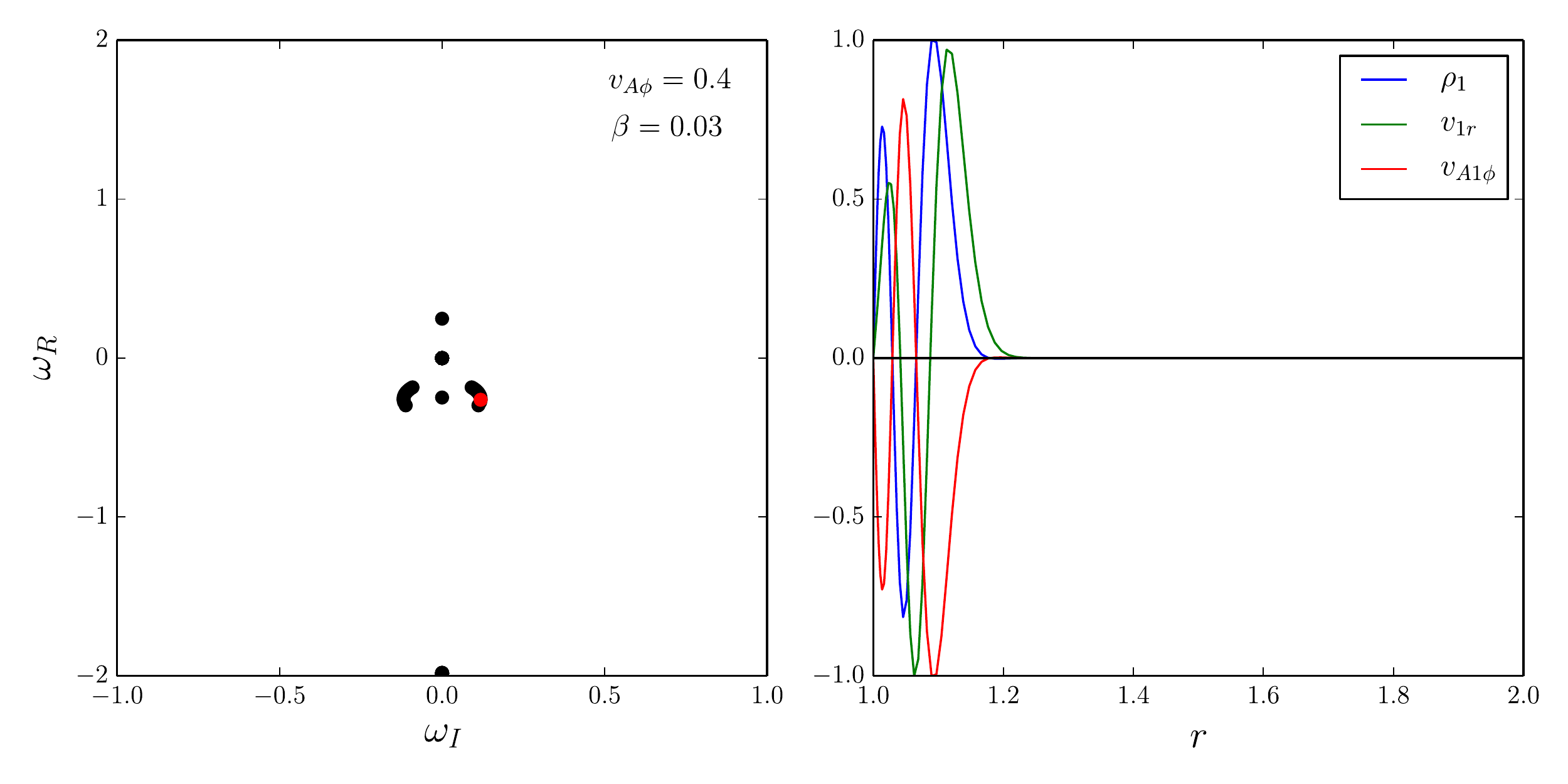}
       \caption{Global properties for $v_{A\phi}=0.4$ computed at $k_z=200$. The left and right panels
       represent the same properties as in Figure \ref{fig_efunckz90}.
       The background accretion flow has $c_s=0.05$ and $v_{Az} = 0.01$,
       and the global problem is solved on a radial grid $r \in [1,5]$. The grid resolution is $N_r=150$ for the left panel
       and $N_r=256$ for the right panel.
       Note that the radial axis in the right panels has been zoomed close to the inner boundary.}
         \label{fig_efunckz200}
\end{figure*}

In the left panels of Figure \ref{fig_efunckz90}, we show the eigenspectrum of all the modes
in the $\omega_R-\omega_I$ plane, for different suprathermal
$v_{A\phi}$ at a fixed $k_z=90 <k_{\rm SI} \approx 93$. The red dot denotes the {\it most} unstable mode
at this $k_z$.
In the right panels of Figure \ref{fig_efunckz90}, we plot three normalized eigenfunctions
corresponding to the most unstable mode at $k_z=90$ (i.e., the red dot in the left panel)
as a function of radius.  These are $\{\rho_1, v_{1r},  v_{A1\phi} \}$,
which are normalized with respect to their respective maximum amplitudes
(for the eigenfunctions of standard weak-field MRI see \citealt{2015MNRAS.453.3257L}).
Figure \ref{fig_efunckz200} is the same as Figure \ref{fig_efunckz90}, except that it is plotted
for $v_{A\phi} =0.4$ at $k_z=200 > k_{\rm SI} \approx 93$ (we do not show the other
cases of Figure \ref{fig_efunckz90}
since there is no instability at $k_z=200$ for $v_{A\phi} < 0.4$). We point out that the instability
represented in this case is the SSMI, while that in the bottommost panels of
Figure \ref{fig_efunckz90} is SHMI
(see \S \ref{subsec_supra} for the definition of $k_{\rm SI}$).
The key features to note from Figures \ref{fig_efunckz90} and
\ref{fig_efunckz200} are as follows:

\begin{enumerate}

\item The eigenspectra tell us that if the
modes lie exactly on the horizontal axis, they are purely unstable (as in the case of
standard weak-field MRI), while if they lie exactly on the
vertical axis then they are purely stable. Any modes that lie in between the two axes are overstable.
Thus, from the left panels of Figures \ref{fig_efunckz90} and
\ref{fig_efunckz200} we clearly see the
overstability of the modes for all the suprathermal cases.
As the field becomes more suprathermal, the family of modes seem to
form more interesting structures in
the $\omega_R-\omega_I$ plane than in the case of standard weak-field MRI.
The growth rates
of {\it all} the unstable modes approach zero as $v_{A\phi} \rightarrow 0.3$, and again
increase for $v_{A\phi}=0.4$, as already seen from Figures \ref{fig_Imw_all} and \ref{fig_Rew_all}.

\item In the right panels, we observe that
all three eigenfunctions
for a given $v_{A\phi}$ look very similar to each other in terms of
their number of nodes as well as radial extent (a node occurs every time
a radial eigenfunction crosses zero).
We observe that the eigenfunctions
of the most unstable mode show different degrees of radial localization for different
$v_{A\phi}$.
We further note that the greater the number of nodes, the less localized the eigenfunctions
tend to be and vice versa. This might be indicative of the potential driving
mechanisms behind the most unstable modes in the suprathermal regime (note, however, that
this is not fully conclusive because of the lack of energetics in our analysis).
The highly radially localized modes
are likely to be shear driven, as the differential rotation dominates
over the restoring magnetic tension force in the inner regions of the disc.
This is the case for standard MRI (see \citealt{2015MNRAS.453.3257L})
and also seems to be so for lower field strengths like $v_{A\phi}=0.1$.
On the other hand, the radially extended modes could be
possibly driven by radial buoyancy caused by
the magnetic tension force of the suprathermal
toroidal field, which tends to dominate in the outer regions of the disc.

\end{enumerate}



\section{Numerical simulations}
\label{sec_pluto}

In this section, we present the results of a small set of numerical simulations that we
performed using the finite volume code PLUTO \citep{2007ApJS..170..228M},
which solves the fully nonlinear equations of ideal MHD. Our motivation is to
verify that the suppression of MRI
as well as the newly identified suprathermal
instabilities can indeed be recovered from numerical simulations of
strongly magnetized accretion discs.
In this work, we focus on the linear regime only, and defer the nonlinear
evolution to a future publication.

\begin{figure*}
\centering
	\includegraphics[width=0.7\textwidth]{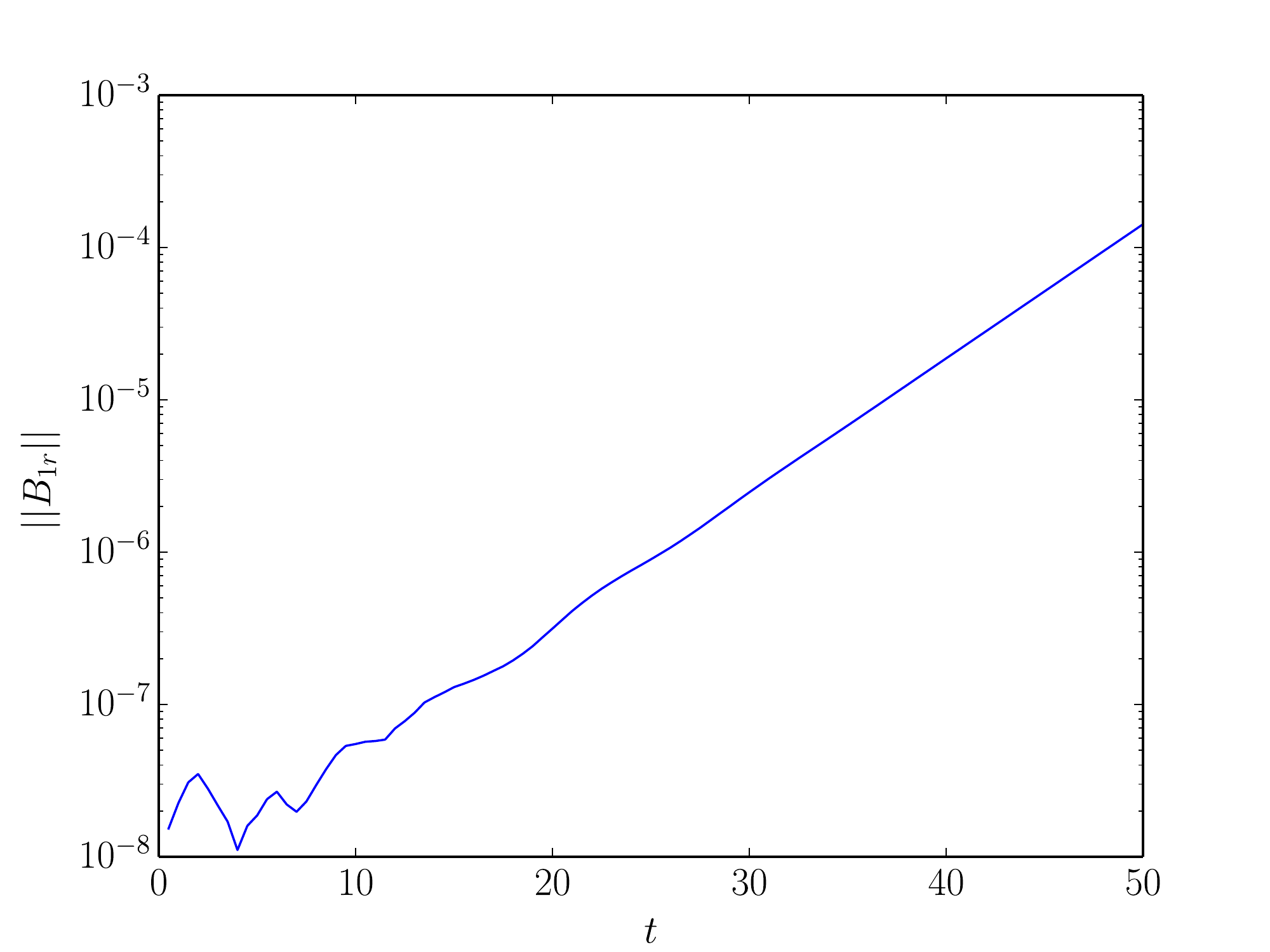}
    \caption{Numerical simulations in PLUTO --- norm of the Fourier-transformed perturbation of
    the radial magnetic field $\lvert \lvert B_{1r} \rvert \rvert$ as a function of
    time $t$, measured at $r_\mathrm{fid}=1.05$ for $v_{A\phi}=0.1$ and $k_z=31$. 
    Time is in units of $\Omega_{\rm in}^{-1}$
    and the measured growth rate in this case is $\omega_{I\mathrm{max}}=0.19$.
    The background accretion flow has $c_s=0.05$ and $v_{Az} = 0.01$, and the simulations are
    performed on a radial grid $r \in [1,5]$ and a vertical grid $z \in [-0.1,0.1]$, with resolutions
    $N_r=1536$ and $N_z=128$ respectively}. 
\label{fig:modes}
\end{figure*}

\subsection{Numerical set-up}

The numerical set-up we consider
matches as closely as possible that of the global linear eigenvalue
analysis performed above (see \S \ref{sec_global} and \S \ref{sec_global_solns}). We work in a 2.5 dimensional
(axisymmetric) geometry and consider a grid extending on a $(r,z)$ domain with $r\in[1,5]$
and $z\in[-0.1,0.1]$. The radial grid resolution is $N_r=1536$ and the vertical grid 
resolution is $N_z=128$.
The boundary conditions are periodic in the vertical direction and outflow
in the radial direction, i.e. $v_{1r}(r=R_{\rm in}=1)=0$ if $v_{1r}(r=1^+)<0$ and $\partial_r v_{1r}=0$ otherwise.
Note that these radial boundary conditions are different than those used for the global eigenvalue
analysis (see \S \ref{sec_eig_setup}), which were actually reflective in nature.
In spite of this difference, we find that the results
of the numerical simulations match quite well with those of the eigenvalue analysis
(see \S \ref{sec_pluto_results} below), indicating that the instabilities we study in this work
are independent of the radial boundary conditions.
In order to limit numerical diffusion in the presence of suprathermal fields, we use a Roe Riemann solver
to compute the intercell fluxes and a third order Runge-Kutta integration scheme. We use a constrained
transport algorithm to ensure $\bm{\nabla\cdot B}=0$ at machine precision and reconstruct the electromotive
forces using a 2 dimensional Riemann solver based on a 4 state HLL (Harten-Lax-Van Leer)
flux function \citep{2007ApJS..170..228M}.

We use the inner radius ($r=R_{\rm in}=1$) of the disc to be the unit of length, and the inverse of the
Keplerian frequency at the
inner radius ($\Omega_{\rm in}^{-1}$) to be the unit of time in
our set-up (note, however, that the disc does not rotate with the Keplerian frequency at the inner radius
due to the non-zero background toroidal field).
We initialize our simulation with the equilibrium rotation profile given
by equation (\ref{eq_radeqbm}), with a uniform mean toroidal and
vertical magnetic field. The flow is assumed to be globally isothermal
and {has a uniform background density $\rho_0=1$.
The simulations presented in this section have $v_{Az}=0.01$ and $c_s=0.05$. The toroidal Alfv\'{e}n
velocity $v_{A\phi}$
is varied from 0 to 0.6. Finally, we add a global perturbation to the flow with 
$v_{1r}=10^{-7}c_s \cos(20 r)\cos(k_z z)$,
which ensures that only the desired vertical wavenumber is initialized in the numerical setup 
(this in turn minimizes contamination due to numerical noise from other scales).

\subsection{Linear evolution}
\label{sec_pluto_results}

We follow the growth of the various instabilities present in this numerical set-up by performing a Fourier
transform in the $z$ direction of the flow at each timestep. In principle, each linearly unstable mode
should grow independently and a projection of the flow on each radial linear eigenmode is required to clearly
identify the growth rate of each mode. Instead of this, we focus here on the {\it most} unstable eigenmode for each
$k_z$. This simplifies the characterization of the instabilities since we can then choose one fiducial
radius $r_\mathrm{fid}$ and measure the growth of the norm of the perturbation at this particular radius.
An example of such a measure is given in Figure \ref{fig:modes},
which shows the norm or, the absolute value, of the Fourier-transformed, perturbed radial magnetic field
at a given $k_z=31$, i.e.,
$\lvert \lvert B_{1r} \rvert \rvert$ as a function of time, for $v_{A\phi}=0.1$. 
The maximum linear growth rate $\omega_{I \rm max}$ that we obtain for this case is $0.19$, 
which agrees quite well with the global eigenvalue analysis, 
as can be seen when compared
with the topmost left panel of Figure \ref{fig_Imw_all} (exact errors given below).
In principle, one could choose any perturbed velocity or magnetic field component,
and we have checked that the growth rates do not depend on this choice. Note also that in principle, one can choose
any $r_{\rm fid}$. However, choosing a large $r_{\rm fid}$ implies that one has to wait for a
long time before the non-growing perturbations have decayed and the most unstable mode shows up.
This is because decaying perturbations typically decay on a timescale $\sim\Omega^{-1}$. Therefore,
it is preferable to choose a $r_{\rm fid}$ close to the inner boundary, but not too close to avoid
potential effects from the boundary conditions. We have chosen $r_\mathrm{fid}=1.05$, which corresponds
to 10 grid cells from the inner radial boundary and is sufficient to avoid artifacts from
the boundary conditions (using $r_\mathrm{fid}=1.1$ did not make any difference to the results).

\begin{figure*}
\centering
	\includegraphics[width=0.7\textwidth]{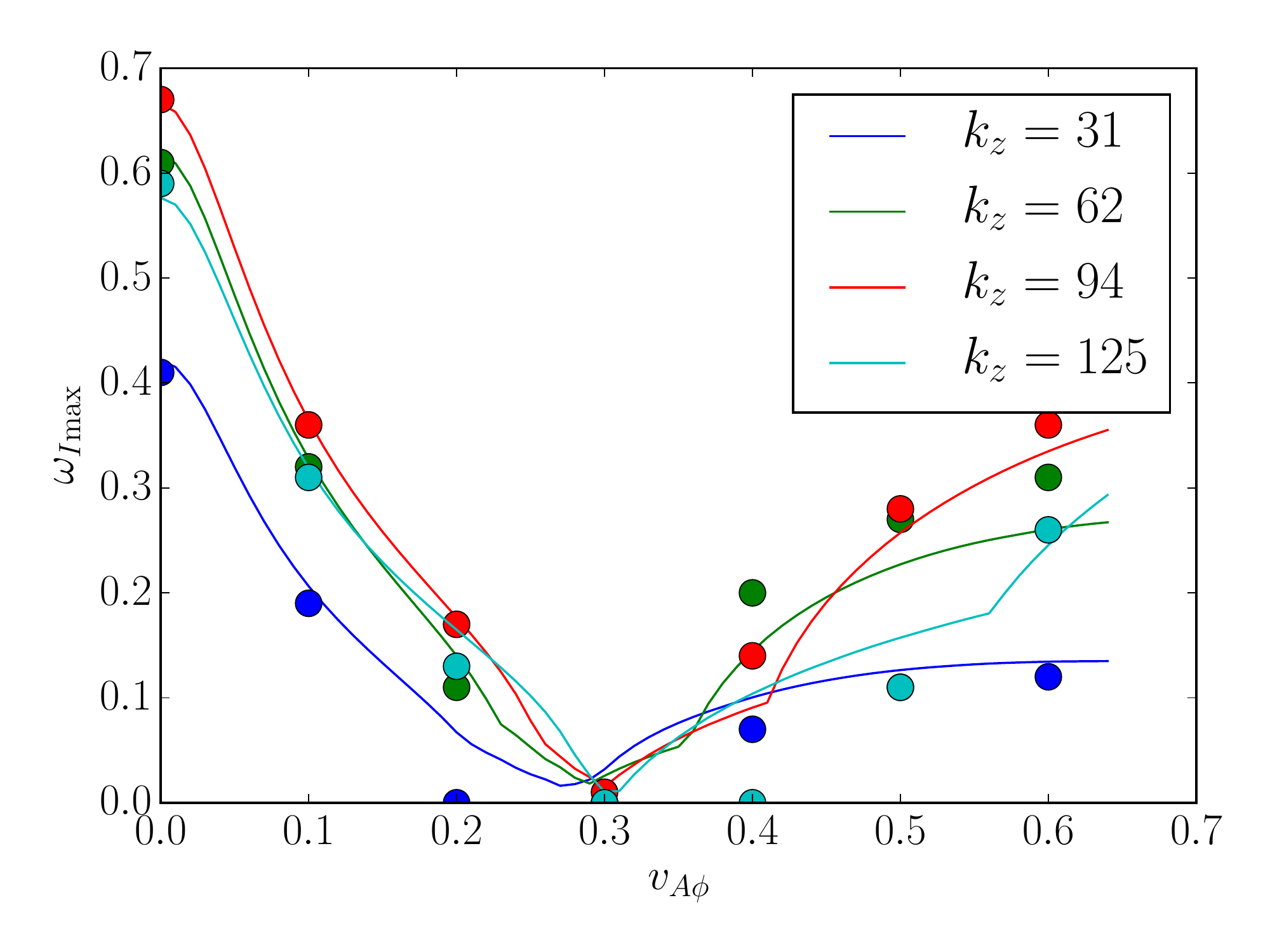}
    \caption{
    Measured maximum linear growth rates $\omega_{I \rm max}$ as a function $v_{A\phi}$ from
   numerical simulations in PLUTO (filled dots) and predicted growth rates from the global eigenvalue
    analysis (lines), for the most unstable mode at $k_z=31,62,94$ and $125$ (represented by different colors).
       The background accretion flow has $c_s=0.05$ and $v_{Az} = 0.01$, and the simulations are
       performed on a radial grid $r \in [1,5]$ and a vertical grid $z \in [-0.1,0.1]$, with resolutions
    $N_r=1536$ and $N_z=128$ respectively}.
    \label{fig:growth}
\end{figure*}

We have performed several simulations with varying $v_{A\phi}$ to measure the growth
rates of the most unstable modes 
at $k_z=31,62,94$ and $125$. Note that because of the vertical 
extension of the domain, we can only probe $k_z$ in multiples of $10\pi$.
The maximum growth rates obtained from PLUTO are shown in
Figure \ref{fig:growth} together with the predicted maximum growth rates
from the global eigenvalue analysis.
Figure \ref{fig:growth} can also be compared with the
left panels of Figure \ref{fig_Imw_all}. We recover the suppression of the MRI growth rate at
$v_{A\phi}=0.3$, as predicted by PP05.
Above $v_{A\phi}=0.3$, we also see the appearance of the new instabilities, namely, SHMI
(see, e.g., the non-zero $\omega_{I \rm max}$ for $k_z=31, 62 < k_{\rm SI}$ for $v_{A\phi}=0.4$)
and SSMI (see, e.g., the non-zero $\omega_{I \rm max}$ for
$k_z=125 > k_{\rm SI}$ for $v_{A\phi}=0.4$, recalling from the eigenvalue analysis that
$k_{\rm SI} \approx 93$ for this field strength). 
The lowest dispersion ($\sigma$) between the numerical and theoretical growth rates
is obtained for $v_{A\phi}=0.1$, with an average dispersion on the four largest modes being
$\sigma=0.01$ (i.e less than $2\%$ in relative error), while the highest error is obtained 
for $v_{A\phi}=0.4$, with $\sigma=0.07$.
Larger discrepancies are observed when small scale modes grow much faster than large scale ones,
in which case our Fourier transform method gets contaminated by the fast-growing modes at small scales. 
This contamination becomes especially important when the magnetic field is highly suprathermal, i.e., 
$v_{A\phi}>0.3$, since the small-scale SSMI modes can have larger growth rates compared to 
the large-scale SHMI modes.




\section{Summary and conclusions}
\label{sec_conc}

Below we summarize the main findings of our analysis:

\begin{itemize}

\item

We have performed a detailed, global eigenvalue analysis of the linearized,
axisymmetric, MHD equations of a differentially rotating fluid
in cylindrical geometry, in order to carry out a
stability analysis of strongly magnetized accretion flows.
We confirm that  MRI growth rates
tend to get highly suppressed in the presence of a sufficiently
suprathermal toroidal magnetic field, when the geometric curvature effects as well as
compressibility of the flow are taken into account.
The current work hence validates one of the main claims of PP05, who performed a
linear stability analysis of a similar accretion flow but under the local approximation.
However, there are some differences between the outcomes of the global
and local analyses, which we discuss below. Note that both the current work and PP05
neglect the effects of non-axisymmetry and spatial gradients in the background variables.

\item We have recovered in our global calculations
that when a limiting toroidal Alfv\'{e}n velocity is reached, the MRI
growth rate starts to decline sharply. This limit is given by
the square root of the product of the Keplerian velocity at
the inner radius and the sound speed, in agreement with the PP05
prediction. However, unlike PP05, we do not observe a complete stabilization
at low vertical wavenumbers at this limit, but instead recover a low growth rate tail extending
all the way to $k_z=0$.
When a {\it second} limit in the toroidal Alfv\'{e}n velocity is reached,
which is a factor $\sqrt{2}$ higher than the previous one, the MRI growth rate drops to
only a few percent of the maximum value for standard MRI (which has zero or very weak
background toroidal field), as predicted by PP05.
However, we find that as long as MRI operates
in the global analysis, it extends
across the entire range of allowed vertical wavenumbers
satisfying the classic Balbus-Hawley criterion of $k_z<k_{\rm MRI}$,
where $(k_{\rm MRI} v_{ A z})^2 \approx - \partial \Omega^2/\partial \ln r$.
This is contrary to PP05, who predicted that the MRI stabilization gradually progresses
from low to higher vertical wavenumbers as the background toroidal field strength increases.

\item We have observed, similar to PP05, the emergence of two new instabilities
in the suprathermal regime beyond the second PP05 limit in our global calculations. However,
the global properties of these instabilities are somewhat different from the local prediction.
These two instabilities are, namely, the suprathermal slow mode instability or SSMI 
 and the suprathermal hybrid mode 
instability or SHMI. We have established that SSMI is a slow mode instability like
MRI, while SHMI is a hybrid mode that results from a destabilized slow-Alfv\'{e}n mode coupling and,
hence, the nomenclature.
In the PP05 case, SSMI and SHMI operate in well-separated vertical wavenumber regimes.
For the same toroidal Alfv\'{e}n velocity, SHMI appears in a {\it limited}
wavenumber range $0 <k_z<k_{c1}$,
while SSMI appears at $k_z>k_{c1}$.
The global analysis exhibits two majors differences in this respect.
First, SHMI in the global case extends all the way to $k_z=0$, unlike in the
PP05 case. Second, there is an overlap in the growth rate of the
{\it most} unstable modes of SHMI
and SSMI at a new critical wavenumber $k_{\rm SI}$ that emerges
in the global analysis. Thus, the most unstable modes of SSMI arise before those of
SHMI can stabilize completely in the $\omega_I-k_z$ space.
There is, however, a sharp discontinuity in the phase velocity of the most unstable modes of these two
instabilities at $k_{\rm SI}$ in the $\omega_R-k_z$ space. This, together with the fact that
SHMI and SSMI have different
physical origins, establish them to be two distinct modes in the global case.
We have further
verified that SSMI stabilizes at a small but finite lengthscale in accordance
with the PP05 stability criteria.
Thus, in the global
analysis, the most unstable modes of
SHMI occupy the range  $0 \leq k_z \leq k_{c1}$, whereas those of SSMI occupy the
whole range  $k_{\rm SI} \leq k_z \leq k_{c2}$, where $k_{\rm SI}$ marks {\it both}
the maximum allowed wavenumber for the {\it most} unstable modes of SHMI as well as the
minimum onset wavenumber for those of SSMI.
The growth rates and the wavenumber
occupancies of both instabilities were seen to increase with an increase in the
background toroidal field strength.
The coexistence of SHMI and SSMI, for the same suprathermal toroidal field,
leads to the interesting
question of whether the SHMI could feed energy into the
SSMI modes and vice-versa in the non-linear regime.

\item On the analytical front, we have self-consistently derived a generic, local dispersion relation,
by using a physically motivated WKB formalism.
A local dispersion relation is useful in
not only providing analytical insight into the problem but also for comparing with
the global calculations. Our dispersion relation includes the effects of
compressibility and magnetic curvature, as well as non-axisymmetry and background radial gradients.
We have further considered the effect of magnetic tension in the background equilibrium
flow, in order to self-consistently
take into account the deviation from Keplerian flow at strong magnetic field strengths.

\item We have computed the normalized radial eigenfunctions of the most unstable modes in
the global analysis, which are not obtainable from a local calculation.
Interestingly, we have found that in the presence of a suprathermal field, the eigenfunctions
of the most unstable modes exhibit different radial localizations.
This is in contrast with standard MRI,
where the most unstable modes are {\it all} localized close to the inner
radial boundary, in keeping with MRI as a purely shear driven instability.
There also seems to exist a positive correlation between the number of radial
nodes and the radial extent of the eigenfunctions.
All this might be an indication of the possible driving mechanisms
for the different suprathermal instabilities.
The modes that are localized close to
the inner radial boundary are likely to be shear driven, whereas the modes that
are more radially extended are likely to be driven by radial buoyancy, which is
generated from the magnetic tension force of the suprathermal toroidal field.

\item We have studied the eigenspectra of the large family of global modes for
different suprathermal fields. They display complex structures in the $\omega_R-\omega_I$
plane, which establishes that all the instabilities, MRI, SHMI and SSMI,
are in general overstable in the presence of a suprathermal toroidal field.
This is contrary to the standard weak-field MRI, which is purely unstable.

\item  We have also verified the main results of our linear eigenvalue analysis
by performing a small set of numerical simulations using PLUTO. The suppression of
the MRI growth rates is clearly recovered at the critical suprathermal
toroidal field strength mentioned above. For toroidal field strengths beyond this
critical value, the new instabilities SSMI and SHMI are also recovered. 
However, for very high
suprathermal fields, there is greater discrepancy between the growth rates from the simulations 
and the eigenvalue analysis. This might be a result of
the increased numerical diffusion at these field strengths, as well as contamination 
from fast growing modes at small scales.


\end{itemize}

Thus, all the above findings underline the need for a global treatment to accurately capture
the curvature effects due to a suprathermal toroidal field. Note that 
in this work, we have assumed the presence of an already suprathermal field, which 
is motivated by the results of shearing box simulations (see, e.g., 
\citealt{2013ApJ...767...30B,2016MNRAS.460.3488S,2016MNRAS.457..857S}). 
The development of a such a field shows that MRI  can 
overcome buoyancy effects in shearing box simulations.
However, shearing boxes cannot take magnetic curvature effects into consideration, 
which is where the relevance of 
the current work sets in. 
The fact that we observe a bottleneck in the MRI growth 
leads to the important question 
of whether it ultimately limits the creation of highly suprathermal toroidal fields 
in real accretion discs. 
Our analysis can serve as a benchmark for global, vertically stratified simulations, which are 
necessary to truly understand the physics of magnetically dominated systems.
While there have been a few such simulations
of strongly magnetized accretion discs
\citep{2000ApJ...532L..67M,2012ApJ...758..103G,2016MNRAS.459.4397S,2017MNRAS.467.1838F},
more work needs to be done to specifically
address and unravel the nature of instabilities in the strongly
magnetized regime.

\section*{Acknowledgments}

This work was supported in part by NASA Astrophysics
Theory Program grant NNX14AB37G and NSF grant AST-1411879.
UD is immensely grateful to Ben Brown for allocating time on the LCD machines,
where all the calculations
using Dedalus were performed, as well as for useful discussions. UD
specially thanks Evan Anders and Jeff Oishi
for their help in navigating through Dedalus and Vladimir Zhdankin for general
discussions.

\bibliographystyle{mn2e}
\bibliography{ref_pp}

\appendix
\section{Correction for non-adiabatic background}
\label{sec_noad_cor}

In the analysis presented in \S \ref{sec_disp_adiab}, we assumed the background to be adiabatic
and, hence, ignored the non-adiabacity ${\bf S}$ given by equation (\ref{eq_adiab}). Here
we relax this assumption.

If we retrace one step, then the second term on the right hand side of equation (\ref{appro_rmom})
is actually $(\rho_1/\rho_0)\rho_0 r(\Omega^2 - \Omega_K^2)$, where
\begin{equation}
\frac{\rho_1}{\rho_0} = \frac{P_1}{\gamma P_0} -  \frac{i v_{1r}}{m\Omega - \omega} S_r
\label{eq_cor_adiab}
\end{equation}
and
\begin{equation}
S_r = \partial_r \biggl(\ln \frac{P_0^{1/\gamma}}{\rho_0} \biggr) ~.
\label{eq_Sr}
\end{equation}
We included the first but not the second term of equation (\ref{eq_cor_adiab}) in equation (\ref{appro_rmom})
and the following steps.
Making the appropriate corrections in the following steps, the dispersion relation given by
equation (\ref{disp_mcb32c}) is modified in the presence of a non-adiabatic background to
\begin{align}
\biggl[(x \mu^2 - n^2) \biggl( 1 + \frac{l^2}{k_z^2}  \biggr)  &+  xy rS_r +
(2 + y) (1 + \hat{B}_\phi) \biggr] \biggl[n^2 - (1+x) \mu^2 \biggr ] \nonumber \\
&+ 2 (2-q)x (1+x)\mu^2 \tilde{\Omega}^2
+ 4\mu x n \tilde{\Omega} (2 + y) + 2 n^2 q x \tilde{\Omega}^2
+ x \mu^2 (2 + y)^2= 0 ~.
\label{disp_mcb32c_app}
\end{align}

\section{Limiting cases of the local dispersion relation}
\label{sec_limitcases}

Here we show that  the dispersion relation given by equation (\ref{disp_mcb32c_app})
reduces to well-known dispersion relations under various limits, which further validates our analytical
calculations.

\begin{enumerate}

\item {\it Non-rotating, no-gravity limit} \citep{1998ApJ...493..291B}:
We put $\tilde{\Omega} = \tilde{\Omega}_K = 0$ and, hence, $y=0$, which reduces
equation (\ref{disp_mcb32c_app}), on some rearrangement, to
\begin{equation}
\frac{1}{2}(x\mu^2 - n^2) \biggl(1 + \frac{l^2}{k_z^2} \biggr) +
\frac{n^2 - (1-x)\mu^2}{n^2 - (1+x)\mu^2}   + \hat{B}_\phi = 0 ~.
\end{equation}
This is identical to the dispersion relation given by equation (3.32) of \citet{1998ApJ...493..291B},
which describes current-driven instabilities of a static MHD pinch. \\

\item {\it Axisymmetric, incompressible, weak-$B_{0\phi}$, no-curvature limit} \citep{1998RvMP...70....1B}:
We put $m=0$ and, hence, $n=\eta$ and $\mu = \tilde{\omega}$, in order to get the axisymmetric dispersion
relation from equation (\ref{disp_mcb32c_app}).
%
%
In addition, we put $\hat{B}_\phi=0$ and
take $x \rightarrow \infty$, $\eta \propto \sqrt{x}$ and everything else finite, which
reduces equation (\ref{disp_mcb32c_app}) after some rearrangement to
\begin{align}
 \biggl(\tilde{\omega}^2 - \frac{\eta^2}{x}\biggr) \biggl( 1 + \frac{l^2}{k_z^2}  \biggr) + y rS_r
 - \frac{2 \tilde{\Omega}^2 \biggl[(2-q)\tilde{\omega}^2
 + q\frac{\eta^2}{x}  \biggr] }{\biggl(\tilde{\omega}^2 -\frac{\eta^2}{x}\biggr)} =0 ~.
\label{disp_mcb36}
\end{align}
Now, in order to compare equation (\ref{disp_mcb36}) with \citet{1998RvMP...70....1B}, we have to
make it dimensionful. Recalling the definitions from Table \ref{tab_notation} and
equation (\ref{eq_radeqbm}), and by multiplying equation
(\ref{disp_mcb36}) throughout by $c_s^4/r^4$, we obtain after some rearrangement 
\begin{align}
 \biggl(\omega^2 - k_z^2 v_{Az}^2\biggr)^2 \biggl( 1 + \frac{l^2}{k_z^2}  \biggr)
 +  \biggl[S_r \frac{1}{\rho_0} \partial_r P_0
 +  \frac{1}{r^3} \partial_r (r^4 \Omega^2) \biggr] (\omega^2 - k_z^2 v_{Az}^2)
 - 4 \Omega^2 (k_z^2 v_{Az}^2)  =0 ~.
\label{disp_mcb36_dim}
\end{align}
We can hence show that equation (\ref{disp_mcb36_dim}) is equivalent to equation
(125) of \citet{1998RvMP...70....1B},
after correcting the typo in the latter
(i.e., ${\cal D}(R^4 \Omega^4) \rightarrow {\cal D}(R^4 \Omega^2)$) and identifying the following:
\begin{align}
(\omega^2 - k_z^2 v_{Az}^2) &\equiv \varpi ~~;~~ k_z^2 v_{Az}^2 \equiv ({\bf k} \cdot {\bf u_A})^2 ~~;~~
\biggl(1 + \frac{l^2}{k_z^2} \biggr) \equiv \frac{k^2}{k_z^2} ~~;~~ r \equiv R ~~;
\nonumber \\ \partial_r &\equiv {\cal D} ~~;~~
P_0 \equiv P ~~;~~ \rho_0 \equiv \rho ~~;~~
S_r \equiv  -\frac{1}{\gamma} {\cal D} \biggl(\frac{\ln P}{\rho^\gamma}\biggr) ~~.
\end{align} \\

\item {\it No poloidal field limit} (TP96):
We first put $\eta=0$ and, hence, $n = m$ in equation (\ref{disp_mcb32c_app}),
which on expanding and dividing throughout by $x(x+1)k^2/k_z^2$,
where $k^2 = l^2 + k_z^2$, yields
\begin{align}
 \mu^4 +
 \biggl[ -\frac{m^2}{x} \frac{(1+2x)}{(1+x)} + 2 q\frac{k_z^2}{k^2} \tilde{\Omega}^2
 - 4 \frac{k_z^2}{k^2}\tilde{\Omega}^2     +  \frac{k_z^2}{k^2} y rS_r  &-
 \frac{k_z^2}{k^2}\frac{(2+y)^2}{(1+x)}  + \frac{k_z^2}{k^2}\frac{(2+y)(1+ \hat{B}_\phi)}{x}   \biggr] \mu^2
 - 4 \frac{k_z^2}{k^2} \frac{(2 + y)}{(1+x)} m \tilde{\Omega} \mu  \nonumber \\ &-
 \frac{m^2}{(1+x)} \frac{k_z^2}{k^2} \biggl[ - \frac{k^2}{k_z^2} \frac{m^2}{x} +  yrS_r
 + 2q \tilde{\Omega}^2 + \frac{(2+y)(1 + \hat{B}_\phi)}{x}  \biggr]  = 0
 \label{eq_ourtp}
\end{align}
or
\begin{equation}
\mu^4 + \tilde{\alpha} \mu^2 - \tilde{\beta} \mu + \tilde{\delta} = 0 ~.
\label{eq_ourtp_equiv}
\end{equation}
The corresponding dispersion relation (dimensionful) from
TP96 is given by their equation (33) as
\begin{equation}
\bar{\sigma}^4 + \alpha_{TP} \bar{\sigma}^2 + \beta_{TP} \bar{\sigma} + \delta_{TP} = 0 ~,
\label{eq_tp33}
\end{equation}
where $\bar{\sigma} = \sigma_{TP} + m\Omega$ such that the time dependence of the perturbations
are of the form $\exp(i \sigma_{TP} t)$; $\alpha_{TP}, \beta_{TP}, \delta_{TP}$ are
given by equations (29), (30) and (34)
of TP96 respectively. We next show that these two equations (\ref{eq_ourtp_equiv}) and (\ref{eq_tp33})
are in fact identical when we set the vertical background gradients in equation (\ref{eq_tp33}) to be zero.

In order to compare the two dispersion relations, we have to first
nondimensionalize equation (\ref{eq_tp33})
according to our prescription such that
\begin{equation}
\mu \equiv -\bar{\sigma} \frac{r}{c_s} ~~;~~ \tilde{\alpha} \equiv \alpha_{TP} \frac{r^2}{c_s^2}
~~;~~ \tilde{\beta} \equiv \beta_{TP} \frac{r^3}{c_s^3} ~~;~~ \tilde{\delta} \equiv \delta_{TP} \frac{r^4}{c_s^4} ~.
\end{equation}
%
First, we consider the coefficient of the quadratic terms in the two dispersion relations.
Equation (29) of TP96
reduces, in terms of our notation (see Table \ref{tab_notation}) to
\begin{equation}
\alpha_{TP} \frac{r^2}{c_s^2} =  -\frac{m^2}{x} \frac{(1+2x)}{(1+x)} +  2 q\frac{k_z^2}{k^2} \tilde{\Omega}^2
 - 4 \frac{k_z^2}{k^2}\tilde{\Omega}^2  + \biggl(\frac{x}{1+x} \biggr) \frac{k_z^2}{k^2} \biggl(y - \frac{2}{x} \biggr)^2
 - \frac{k_z^2}{k^2} y \biggl(\frac{r \partial_r \rho_0}{\rho_0} \biggr) + \frac{2}{x}\frac{k_z^2}{k^2} (\hat{B}_\phi -1) ~.
 \label{eq_alphatp}
\end{equation}
We see that the first three terms in the above equation match exactly with the first three terms
in the coefficient of $\mu^2$ (i.e., $\tilde{\alpha}$) in equation (\ref{eq_ourtp}).
Canceling the common factor of $k_z^2/k^2$, we therefore
need to show that the remaining terms between equations (\ref{eq_ourtp}) and (\ref{eq_alphatp}) also agree, namely
\begin{equation}
y rS_r - \frac{(2+y)^2}{(1+x)}  + \frac{(2+y)(1+ \hat{B}_\phi)}{x} =
\biggl(\frac{x}{1+x} \biggr) \biggl(y - \frac{2}{x} \biggr)^2
 -  y \biggl(\frac{r \partial_r \rho_0}{\rho_0} \biggr) + \frac{2}{x} (\hat{B}_\phi -1) ~.
 \label{eq_mu2_extra}
\end{equation}
Using equations (\ref{eq_radeqbm}) and (\ref{eq_Sr}) we can write
\begin{equation}
-\frac{r \partial_r \rho_0}{\rho_0} = r S_r  - r\frac{\partial_r P_0}{\gamma P_0} = r S_r - y + \frac{1 +\hat{B}_\phi}{x} ~.
\label{eq_Sr_mod}
\end{equation}
The above, when replaced on the right hand side of equation (\ref{eq_mu2_extra}),
makes it identical to the left hand side. This establishes that the coefficient of $\mu^2$ in our
dispersion relation (equation \ref{eq_ourtp}) agrees with that of TP96.

Similarly, we can show the coefficient of the linear terms as well
as the constant terms are identical between the two dispersion relations.
Thus, our dispersion relation given by equation (\ref{disp_mcb32c_app})
is equivalent to that of TP96, when $n=m$ and the background state depends only on $r$.

\end{enumerate}

\label{lastpage}
\end{document}